\documentclass[manuscript,natbib=true]{acmart}

\usepackage{enumitem}
\usepackage{natbib}

\begin{document}
\title[Entities as Retrieval Signals]{Entities as Retrieval Signals: A Systematic Study of Coverage, Supervision, and Evaluation in Entity-Oriented Ranking}
\author{Shubham Chatterjee}
\affiliation{%
  \institution{Missouri University of Science and Technology}
  \city{Rolla}
  \country{USA}
}
\email{shubham.chatterjee@mst.edu}
\begin{abstract}
Entity-oriented retrieval is built on the intuition that documents
relevant to a query should exhibit entities relevant to the user's
information need. Yet current evaluations give conflicting answers about
whether entity signals actually help ranking. We show that the
inconsistency is not a model failure; rather, it is an evaluation failure.

We conduct a systematic study on TREC Robust04, evaluating six neural
entity-oriented rerankers alongside 437 unsupervised entity-oriented
configurations and a BM25 baseline. Across 443 configurations, no system
improves MAP by more than 0.051 over BM25 once entity selection is made
independent of the relevance judgments used for evaluation---despite
the much stronger gains reported when it is not. We show that this gap is
produced almost entirely on the entity side: holding the document ranker
fixed, the choice of entity run moves MAP by roughly 0.4, while
restricting or expanding the document candidate pool moves it by less
than 0.005. The best configuration under independent entity selection
matches the official Robust04 best system and outperforms the majority of
neural rerankers, confirming that the architecture is not the problem.
The bottleneck is the entity channel itself: even under binary-derived
supervision---the strongest signal these benchmarks can produce without
external annotation---entity selection covers only 19.7\% of relevant
documents in the BM25 candidate pool, and no configuration across 193
entity selection strategies simultaneously achieves high coverage and
high discrimination.es simultaneously achieves high
coverage and high discrimination.

To explain this, we distinguish \emph{Conceptual Entity Relevance} (CER)--- whether an entity is semantically related to a query---from
\emph{Observable Entity Relevance} (OER)---whether the presence of that
entity, as detected by the entity linker in this collection, actually
increases the probability that a document is relevant. No supervision
strategy evaluated targets OER: binary derivation targets exclusivity,
LLM-generated labels target conceptual relatedness, and corpus-grounded
selectors approximate discriminativeness only indirectly. None accounts
for the linking environment that determines observable
discriminativeness. This
mismatch explains why better supervision does not recover open-world
performance: conceptually correct entity labels are not the same as
observably discriminative ones, and filtering for stronger signals reduces
coverage without recovering discrimination.

The core finding is that entity supervision derived from document relevance judgments and entity supervision constructed independently of them answer different scientific questions. The former asks whether a model can exploit entity evidence when it is aligned with relevance, whereas the latter asks whether the full pipeline improves retrieval when that alignment must be inferred. The field has often reported answers to the first question as though they answered the second. Progress requires collections with aligned
entity-level discriminativeness annotations and evaluation protocols that
report coverage alongside effectiveness. Until then, gains measured under
derived entity supervision should not be read as evidence of end-to-end
effectiveness, and the absence of gains under independent supervision
should not be read as evidence that entity-oriented architectures fail.
\end{abstract}
\maketitle

\section{Introduction}
\label{sec:Introduction}

Entity-oriented retrieval is motivated by a simple intuition: documents
relevant to a query are more likely to exhibit entities that are
themselves relevant to the user's information need, and these entities
can serve as evidence of document relevance. For example, for the query
\textit{Black Bear Attacks}, a document mentioning entities such as
\textit{Parks Highway}, known for bear sightings, provides stronger
evidence than one mentioning peripheral entities such as \textit{National Guard}. The goal is thus to leverage
\emph{query-relevant} entities to identify
relevant documents beyond surface term matching.

This intuition underlies a wide range of entity-oriented ranking models.
EQFE~\cite{dalton2014entity} explicitly states it, identifying
queries-relevant entities and using them as signals of document relevance.
Other approaches adopt the same view implicitly: the Word-Entity
Duet~\cite{xiong2017word}, JointSem~\cite{xiong2017jointsem}, and
EsdRank~\cite{xiong2015esdrank} model entities as intermediate signals
that connect queries and documents, while EDRM~\cite{liu-etal-2018-entity}
incorporates entity semantics into neural ranking. Recent models~\cite{chatterjee2024dreq,chatterjee2025qder} make this bridge explicit by ranking
entities and using them to guide document ranking. Despite their architectural
differences, these approaches share the same premise: entity-level
evidence can improve document retrieval.

This raises a fundamental question: \emph{Can current evaluation settings in
entity-oriented retrieval meaningfully assess the usefulness of entity
signals for document ranking?} In this paper, we argue that they often
cannot. More specifically, can these evaluation settings actually test
the long-standing hypothesis that relevant documents exhibit
entity-level evidence relevant to the query? 

In this paper, we study \emph{document re-ranking} over an initial set of
retrieved candidates. We find that current benchmarks fail to show
consistent gains from entity-based methods once entity selection is made
independent of the relevance judgments used for evaluation, despite the
strong improvements reported when it is not. This creates an empirical
inconsistency: \emph{the same models, over the same candidate pool and
under the same metric, appear effective under one supervision regime and
ineffective under another.}

This inconsistency has recently been observed from the outside.
Boud\`ens et al., 2026~\cite{boudens2026relevance} attempted to reproduce
QDER~\cite{chatterjee2025qder} and reported that its effectiveness
depends on relevance information entering the pipeline through derived
entity supervision. We confirm that diagnosis and quantify it here: on
Robust04, entity features reach 93.8\% of relevant candidates but only
44.6\% of non-relevant ones, because the entity set is derived from the
document judgments themselves. Their study attributes the effect
primarily to filtering of the document candidate pool; we show that the
document pool contributes under $0.005$ MAP and that the entity run
accounts for essentially all of the difference. We note also that their
reproduction annotates Robust04 at roughly a third of our linking density:
29 unique entities per document against our 78.5, and 10{,}645 unique
entities per query candidate pool against our 31{,}346. The underlying
entity vocabularies are comparable in size (557{,}334 against 643{,}271),
so the difference lies in linker permissiveness rather than in
knowledge-base coverage. Since the coverage ceiling is a property of the
linking environment, effectiveness figures are not directly comparable
across annotations, and we report ours in
Section~\ref{subsec:Experimental Setup}.

We show that this discrepancy does not arise from the ranking architecture
itself, but from the interaction between entity signals and the evaluation
setting. Current benchmarks do not provide a setting in which the
usefulness of entity signals can be isolated from how the entity channel
is constructed, filtered, and observed. As a result, evaluation entangles entity signal quality with coverage,
hindering a clear assessment of their contribution. This issue is compounded by the absence of entity-level supervision in standard benchmarks, which forces entity-oriented methods to rely on proxy signals derived from document relevance.

To see why, it is important to separate two choices that current
evaluations conflate. The first concerns the \emph{entity run}: whether
the entity selection supplied to the document ranker is derived from the
relevance judgments of the query being evaluated, or constructed without
access to them. The second concerns the \emph{document pool}: whether
ranking is restricted to candidates containing a selected entity, or
performed over the full set. These are independent, and they are not
comparable in magnitude --- on Robust04 the entity-run choice moves MAP
by roughly $0.4$, the document-pool choice by less than $0.005$.

Under \emph{leaked} entity supervision, the pipeline answers: \emph{if
entity signals aligned with relevance are available, can the model
exploit them?} That is a meaningful question about model capacity. Under
\emph{clean} entity supervision, it answers: \emph{does the full
entity-oriented pipeline improve retrieval when entity selection must be
performed without relevance judgments?} That is a meaningful question
about practical effectiveness. Both are valid, they are not equivalent,
and neither subsumes the other.

This distinction matters because entity-oriented methods can perform well
under one regime and poorly under the other for reasons that have little
to do with the ranking architecture itself. A system may achieve strong
effectiveness under leaked supervision because the entity run encodes the
relevance judgments against which it is scored. Yet the same system may
show little gain under clean supervision if the entities it can select
without those judgments cover only a small fraction of the relevant
documents.

Conversely, broadening the entity channel can reduce the gap between the
two regimes by increasing coverage, but only at
the cost of introducing many weak, generic, or non-discriminative
entities. Failure under clean supervision therefore does not imply that
results measured under leaked supervision were arithmetically wrong, nor
does the latter imply end-to-end usefulness. The two regimes answer
different questions, and current benchmarks do not cleanly reconcile them.

We demonstrate this tension through a systematic study on the TREC
Robust04 collection, evaluating six entity-oriented neural ranking
models---EDRM~\cite{liu-etal-2018-entity}, Word-Entity Duet~\cite{xiong2017word}, EsdRank~\cite{xiong2015esdrank}, EVA~\cite{tran2022dense}, DREQ~\cite{chatterjee2024dreq}, and QDER~\cite{chatterjee2025qder}---alongside
437 unsupervised entity-oriented ranking configurations and a BM25
baseline. Under open-world evaluation over the full BM25 candidate pool,
no configuration improves MAP by more than 0.051 over BM25 (0.34
vs.\ 0.29), despite the much stronger gains reported when entity
selection is derived from the evaluation judgments.
Crucially, this pattern holds across model families. The architecture is
not the primary limitation.

Instead, the bottleneck lies in the entity channel itself. Standard ad hoc
retrieval collections do not provide entity-level relevance judgments, so
entity-oriented methods must rely on proxy supervision to learn which
entities matter for a query. At the same time, entity linking is
incomplete, and many relevant documents do not expose the entities that
would be needed to support them through the entity channel.

These two limitations create a structural dilemma. Restricting entity
selection to cleaner or higher-confidence signals can improve precision,
but it reduces coverage of relevant documents. Broadening entity selection
can improve coverage, but it introduces more noise and weaker signals. As
a result, current benchmarks offer no single evaluation setting in which
the usefulness of entity signals can be assessed independently of their
coverage. Leaked supervision suppresses the coverage problem by
selecting entities the judgments guarantee are aligned with relevance;
clean supervision entangles entity usefulness with coverage. This is the
sense in which evaluation in entity-oriented IR is ill-posed under
current settings.

To make this failure mode explicit, we introduce a diagnostic framework
that separates two notions of entity usefulness under current benchmark
constraints. Rather than improving entity linking or assuming access to
entity-level supervision, we ask a more operational question: does
observing an entity in a document increase the likelihood that the
document is relevant?
We distinguish between \emph{Conceptual Entity Relevance} (CER), which
captures whether an entity is semantically related to the query, and
\emph{Observable Entity Relevance} (OER), which measures whether that
entity provides discriminative evidence of relevance in the collection.
This distinction aligns usefulness with what can be observed, explaining
why better-looking supervision does not recover open-world performance:
many conceptually relevant entities are not discriminative, while
filtering for stronger signals reduces coverage.

\medskip
\noindent
\textbf{Empirically, this distinction appears as a consistent failure
mode.} Across 443 entity-oriented re-ranking configurations, no system improves
MAP by more than 0.05 over BM25 on the full candidate set, despite
matching the official Robust04 best system, indicating that the
limitation is not architectural. 

Performance varies little with the document pool (under $0.005$ MAP), but
shifts substantially with the entity run (up to $0.40$ MAP). Even under
binary-derived supervision --- the strongest signal the benchmark can
produce without external annotation --- the entity channel reaches only
19.7\% of relevant documents in the BM25 pool.
Together, these results show that entity-oriented effectiveness is
confounded with coverage: outcomes are determined almost entirely by how
the entity run is constructed, and leaked and clean supervision capture
distinct aspects of entity usefulness.

\medskip
\noindent
\textbf{Contributions.} We make the following contributions:
\begin{itemize}
    \item We show that leaked and clean entity supervision answer
    different scientific questions, and that gains reported under the
    former do not transfer to the latter across any of the 443
    configurations we evaluate --- including strong neural models that
    match or outperform the best available neural rerankers under clean
    supervision. We further show that this gap is not produced by
    restricting the evaluation pool, as has been suggested: the document
    pool contributes under $0.005$ MAP.

    \item We demonstrate that the open-world collapse is not
    architectural but structural: it originates on the entity side, is
    explained by a coverage ceiling imposed by the linking environment,
    and persists across all supervised and unsupervised configurations
    regardless of model family or supervision strategy.

    \item We introduce the CER--OER distinction between conceptual entity relevance and observable entity relevance, and show that this mismatch is the root cause of the supervision--coverage dilemma. None of the supervision strategies evaluated targets OER: binary derivation targets exclusivity, LLM labels target conceptual relatedness, and corpus-grounded selectors approximate discriminativeness only indirectly. Refining these strategies cannot recover the OER signal that the benchmark does not annotate.

    \item We show empirically that the coverage--discrimination tradeoff
    is inescapable: across 193 entity selection configurations, no
    strategy simultaneously achieves high relevant-document coverage and
    high discriminative precision, and directly enforcing OER alignment
    via post-hoc filtering improves signal quality but reduces coverage
    to the point that ranking falls below BM25.

    \item We provide a diagnostic framework --- covering leaked vs.\
    clean entity supervision, document-pool effects, OER-based
    entity-level metrics, and
\end{itemize}

\section{Framework and Experimental Setup}
\label{sec:Framework and Experimental Setup}

\subsection{Problem Setting}
\label{subsec:Problem Setting}

Let $\mathcal{D}$ be a document collection and $Q$ a set of queries
with relevance judgments $\mathcal{R} \subseteq Q \times \mathcal{D}$.
An entity linker $\Lambda$ maps each document $d \in \mathcal{D}$ to a
set of entities $\Lambda(d) \subseteq \mathcal{E}$, where
$\mathcal{E}$ is a knowledge-base entity vocabulary. We study
\emph{entity-oriented document re-ranking}: given a query $q$ and an
initial BM25 candidate set $\mathcal{D}_{\text{cand}}(q)$, a reranker
produces a refined ranking over those candidates.

Entity-oriented rerankers augment ranking with linked-entity signals.
These may be incorporated by encoding entities in query and document
representations~\cite{liu-etal-2018-entity,xiong2017word}, jointly
modeling linking and ranking~\cite{xiong2017jointsem}, or ranking
query-specific entities to guide document
scoring~\cite{chatterjee2024dreq,chatterjee2025qder}. Related work also
injects entity information into dense~\cite{tran2022dense} and sparse
retrieval~\cite{nguyen-etal-2024-dyvo}. We treat all of these as
instances of a common step: selecting which entity signals influence
document ranking.
Formally, we decompose this process into two steps. First, an entity
selection component associates query $q$ with a ranked list of candidate
entities $\hat{E}_q = [e_1, e_2, \ldots]$, where each
$e_i \in \mathcal{E}$. When explicit scores are available, selection is
defined by the top-$k$ prefix, denoted $\hat{E}_q^k$.

This abstraction unifies several model classes. In EDRM~\cite{liu-etal-2018-entity},
$\hat{E}_q$ is simply the set of linked entities, with no separate
ranking stage or explicit top-$k$ truncation. In attention-based models
such as Word-Entity Duet~\cite{xiong2017word}, linked entities are
retained but weighted, which can be viewed as soft selection over
$\hat{E}_q$. Retrieval-based approaches such as
EsdRank~\cite{xiong2015esdrank} construct $\hat{E}_q$ by retrieving
related objects from external resources. More recent methods make this
step explicit by learning an entity scoring function and selecting the
top-$k$ for downstream use~\cite{chatterjee2024dreq,chatterjee2025qder}.

For query $q$, let $\mathcal{R}_q$ denote the relevant documents and
$\overline{\mathcal{R}_q}$ the non-relevant documents in
$\mathcal{D}_{\text{cand}}(q)$. Let $df_{\text{rel}}(e,q)$ and
$df_{\text{nonrel}}(e,q)$ denote the numbers of relevant and
non-relevant candidate documents containing entity $e$, and let
$df_{\text{cand}}(e,q) = df_{\text{rel}}(e,q) + df_{\text{nonrel}}(e,q)$
be its total candidate-pool frequency. All quantities are computed over
$\mathcal{D}_{\text{cand}}(q)$.

\subsection{Evaluation Settings}
\label{subsec:Evaluation Settings}

Entity-oriented re-ranking pipelines admit two distinct evaluation
choices, and conflating them has caused persistent confusion. The first
concerns the \emph{entity run}: whether the entity selection fed to the
document ranker is derived from the same relevance judgments used for
final evaluation, or is constructed without access to them. The second
concerns the \emph{document pool}: whether ranking is restricted to
candidates that contain a selected entity, or performed over the full
candidate set. These are independent, and as we show in
Section~\ref{subsec:entity_side_bottleneck} they are not
comparable in magnitude: the entity-run choice moves MAP by roughly
$0.4$, the document-pool choice by less than $0.005$. We define both,
but the entity run is the axis that matters.

\subsubsection{Leaked and Clean Entity Supervision}
\label{subsubsec:entity-run-setting}

Under \emph{leaked} entity supervision, the entity run supplied to the
document ranker at test time is produced by a ranker trained on entity
labels derived from the test queries' own document judgments via
Equation~\ref{eq:binary_derivation}, and restricted to the entity pool
those judgments induce. Under \emph{clean} entity supervision, the entity
run is constructed without reference to the relevance judgments of the
query being evaluated: the ranker scores the full unfiltered entity pool
induced by $\bigcup_{d \in \mathcal{D}_{\text{cand}}(q)} \Lambda(d)$, or
the selection is produced by a method that uses no relevance labels at
all.

This distinction is the primary determinant of measured effectiveness in
entity-oriented ranking, and it is orthogonal to the document pool.
Published results for QDER and DREQ on Robust04 were computed with
\texttt{trec\_eval -c} over the full 1000-document candidate set for
every topic, with unrestricted recall denominators --- that is, under
\emph{open-world document pools}. Their high effectiveness is therefore
not an artifact of pool restriction; it follows from leaked entity
supervision.

\subsubsection{Conditional Evaluation}
\label{subsubsec:conditional-evaluation}

Under conditional evaluation, ranking is restricted to candidate
documents that contain at least one selected query entity:
\begin{equation}
\mathcal{D}^{\text{cond}}_q
=
\left\{
d \in \mathcal{D}_{\text{cand}}(q)
\;\middle|\;
\Lambda(d) \cap \hat{E}^k_q \neq \emptyset
\right\}.
\label{eq:conditional}
\end{equation}
Metrics are computed over $\mathcal{D}^{\text{cond}}_q$, with recall
denominators restricted to
$\mathcal{R}_q \cap \mathcal{D}^{\text{cond}}_q$.

We note that pool restriction \emph{without} a corresponding denominator
restriction has the opposite effect: evaluated with \texttt{trec\_eval -c}
against the full judgment set, a filtered run is penalised for the
relevant documents it discards. On Robust04 (title), the QDER stage run
restricted in this way scores MAP $0.5667$, against $0.6082$ for the same
model over the full 1000-document pool. Filtering and denominator
restriction therefore move scores in opposite directions, and reports
that do not state which was applied are not interpretable.

\subsubsection{Open-world Evaluation}
\label{subsubsec:open-world-evaluation}

Under open-world evaluation, ranking is performed over the full
candidate set regardless of entity presence:
\begin{equation}
\mathcal{D}^{\text{ow}}_q = \mathcal{D}_{\text{cand}}(q).
\label{eq:openworld}
\end{equation}
Metrics are computed over $\mathcal{D}^{\text{ow}}_q$ against
$\mathcal{R}_q \cap \mathcal{D}^{\text{ow}}_q$. This corresponds to the
standard ad hoc setting, where all retrieved candidates must be ranked,
including those for which the entity channel provides little or no
usable signal.

\medskip
\noindent
\textbf{Why the Distinction Matters.}
%
Open-world evaluation measures end-to-end effectiveness over all
candidates, including cases where the entity channel is weak, absent, or
noisy. Conditional evaluation instead isolates the subset of documents
for which the entity channel can contribute. The two settings therefore
answer different questions and should not be treated as interchangeable.

If many relevant documents lack linked instances of the selected
entities, conditional evaluation with restricted denominators can
overstate the usefulness of entity signals by excluding documents the
entity pipeline cannot reach. We formalize this through the coverage
metrics in
Section~\ref{subsec:Coverage and Discrimination Metrics}.

This coverage argument explains why entity runs built \emph{without}
access to relevance judgments fail to exceed BM25 by a meaningful margin
under open-world evaluation. It does not explain the high effectiveness
reported for leaked entity supervision, which is measured over the full
candidate pool with unrestricted denominators. The two phenomena have
distinct causes and we treat them separately throughout.

\subsection{Coverage and Discrimination Metrics}
\label{subsec:Coverage and Discrimination Metrics}

To quantify when entity signals can meaningfully contribute to
document ranking, we introduce a set of diagnostic metrics that
characterize the reachability and selectivity of entity evidence
within the BM25 candidate pool. These are not standard IR
effectiveness measures; rather, we designed them to capture
structural properties of the query--collection--linker pipeline
independent of ranking quality.

Specifically, we define two complementary coverage metrics:
relevant coverage and non-relevant coverage. From these, we
derive a discrimination statistic that quantifies how selectively
the entity signal covers relevant documents relative to
non-relevant ones.

\subsubsection{Relevant/Non-Relevant Coverage}
\label{subsubsec:relcov}
The relevant coverage at cutoff $k$ for query $q$ is the fraction
of relevant candidate documents that contain at least one top-$k$
entity:
\begin{equation}
\text{RelCov}(q, k) = \frac{\left|\left\{ d \in \mathcal{R}_q \cap
\mathcal{D}_{\text{cand}}(q) \;\middle|\; \Lambda(d) \cap
\hat{E}^k_q \neq \emptyset \right\}\right|}{|\mathcal{R}_q \cap
\mathcal{D}_{\text{cand}}(q)|}
\label{eq:relcov}
\end{equation}
$\text{RelCov}(q, k)$ measures the reachability of relevant
documents through the entity channel. A relevant document with
$\Lambda(d) \cap \hat{E}^k_q = \emptyset$ is structurally
invisible to any entity-oriented pipeline regardless of the
quality of the downstream document ranker.
$\text{NonRelCov}(q, k)$ is defined analogously over non-relevant
documents, measuring the fraction of non-relevant candidates that
also contain a top-$k$ entity.

$\text{RelCov}(q, k)$ is recall-like but independent of ranking:
it measures the fraction of relevant documents that are reachable
through the entity channel, rather than those actually retrieved. $\text{NonRelCov}(q, k)$ quantifies the spread of the same entity
signal over non-relevant documents, analogous to a false activation
rate rather than a ranking error.

\subsubsection{Discrimination Ratio}
\label{subsubsec:discratio}
The discrimination ratio at cutoff $k$ for query $q$ is:
\begin{equation}
\text{DiscRatio}(q, k) = \frac{\text{RelCov}(q, k)}
{\text{NonRelCov}(q, k) + \epsilon}
\label{eq:discratio}
\end{equation}
where $\epsilon$ is a small smoothing constant. A high
discrimination ratio indicates that the selected entities appear
substantially more often in relevant than non-relevant documents;
a ratio near 1.0 indicates the entity signal is non-discriminative.
We report $\text{RelCov}@k$, $\text{NonRelCov}@k$, and
$\text{DiscRatio}@k$ for $k \in \{10, 20, 50\}$ throughout.

Unlike precision, which depends on ranking, the discrimination
ratio measures how selectively the entity signal covers relevant
documents relative to non-relevant ones, independent of ranking.

\subsubsection{Coverage Ceiling}
\label{subsubsec:Coverage Ceiling}

We define the coverage ceiling as $\mathbb{E}_q[\text{RelCov}(q,
k)]$ for a given collection--linker--pipeline combination. This
quantity is a hard upper bound on the recall achievable by any
entity-based pipeline: relevant documents with $\Lambda(d) \cap
\hat{E}^k_q = \emptyset$ cannot be retrieved through the entity
channel regardless of model quality. The coverage ceiling is
jointly determined by the entity linker, the knowledge base, and
the collection --- it is a property of the evaluation environment,
not of any individual model.

Unlike recall, which depends on retrieval performance, the coverage
ceiling is a structural upper bound determined by the entity
pipeline and collection.

\subsection{Derived Entity Supervision}
\label{subsec:Derived Entity Supervision}

Standard ad hoc retrieval collections provide document-level relevance
judgments but not entity-level relevance annotations. In settings such as
TREC Complex Answer Retrieval~\cite{dietz2017trec}, supervision can be
derived from hyperlink structure or document organization, and in TREC
Deep Learning~\cite{craswell2025overview}, document-level relevance is
often inferred from passage-level evidence. No analogous entity-level
annotations exist for collections such as Robust04. For entity-oriented
retrieval, supervision must therefore be derived from document relevance.

Following this paradigm, we derive entity labels with a binary rule:
\begin{equation}
\text{label}(e, q) = \begin{cases}
1 & \text{if } df_{\text{rel}}(e, q) > 0
    \text{ and } df_{\text{nonrel}}(e, q) = 0 \\
0 & \text{if } df_{\text{rel}}(e, q) = 0
    \text{ and } df_{\text{nonrel}}(e, q) > 0 \\
\text{excluded} & \text{if } df_{\text{rel}}(e, q) > 0
    \text{ and } df_{\text{nonrel}}(e, q) > 0
\end{cases}
\label{eq:binary_derivation}
\end{equation}

This partitions the entity space for query $q$ into exclusive positives
$\mathcal{E}_q^+$, exclusive negatives $\mathcal{E}_q^-$, and common
entities $\mathcal{E}_q^0$, which are excluded from supervision. The
intuition is straightforward: entities observed only in relevant
documents are more plausible query-specific signals, whereas entities
observed in both relevant and non-relevant documents provide ambiguous
evidence. Rather than resolving that ambiguity with an additional
heuristic, we omit such entities from training. This is broadly
consistent with prior work that suppresses noisy or weak entity
signals~\cite{dalton2014entity,xiong2017word,xiong2017jointsem,tran2022dense},
but here the criterion is defined directly from query-specific relevance
judgments.

This supervision is necessarily only a proxy. Unlike passage-to-document
transfer, where relevance propagates upward along a containment relation,
document-to-entity transfer requires inferring which entities within a
relevant document are actually useful for the query. That inference is
substantially weaker: an entity may appear in a relevant document without
being important to the query or helpful for distinguishing relevant from
non-relevant documents. Under incomplete entity linking, this creates two
opposing risks: over-including weak entity signals and under-covering
useful ones. We analyze the resulting supervision--coverage trade-off in
Section~\ref{sec:The Supervision--Coverage Dilemma}.

This proxy is not a methodological convenience but a structural
requirement of the benchmark setting. Standard ad hoc collections such
as Robust04 and TREC DL annotate document relevance, not entity
relevance. As a result, deriving entity labels from document relevance
introduces a real circularity: document judgments define which entities
are treated as relevant, and those derived labels are then used to guide
document ranking evaluated against the same document judgments.
Leaked entity supervision hides this circularity: the entity run encodes
the relevance judgments against which the document ranking is scored, so
effectiveness remains high even when every candidate is ranked and every
relevant document counted. Clean entity supervision exposes the
circularity by removing it, and reveals the coverage ceiling that
remains. This is not a flaw of one
particular method, but a consequence of testing the field's central
hypothesis without independent entity-level discriminativeness
annotations. Making that circularity explicit, characterizing where it
yields useful signal and where it breaks down, is central to the
analysis in this paper.

\subsection{Observable Entity Relevance}
\label{subsec:Observable Entity Relevance}

The two failure modes identified above---over-inclusion of generic
entities and under-coverage of truly discriminative ones---stem from a
common mismatch: how entity relevance is defined versus how entity
evidence is actually observed under linking. The supervision strategies
considered in this work assess whether an entity is topically related to
a query. That is a natural notion of relevance, but in a noisy linking
environment it is not the quantity that retrieval depends on.

The standard conceptual view assumes that entities central to a query
will be reliably linked in the documents that discuss them. This
assumption is approximately satisfied in structured collections such as
TREC CAR~\cite{dietz2017trec}, where annotations come directly from
Wikipedia hyperlinks, but it is much weaker in newswire and web
collections such as Robust04 and TREC Web. There, entity linking is
noisy: mentions may be missed, surface forms conflated, and annotations
applied inconsistently across documents. Under these conditions, the key
question is not whether entity $e$ is semantically related to query $q$,
but whether observing $e$ under linking provides evidence that a
document is relevant.

Formally, the operational question is whether
\[
P(e \in \Lambda(d) \mid d \in \mathcal{R}_q)
\]
is meaningfully larger than
\[
P(e \in \Lambda(d) \mid d \notin \mathcal{R}_q).
\]
An entity may be highly relevant in the conceptual sense yet fail this
test: if it is not linked consistently in relevant documents, it
provides little coverage. Conversely, an entity may be topically
plausible but so frequent across the collection that its presence does
not distinguish relevant from non-relevant material. These are the two
failure modes above, viewed through a common lens: conceptual
relatedness and observable discriminative utility are not the same.

This motivates a different criterion. Rather than asking \emph{is this
entity related to the query?}, we ask \emph{does the presence of this
entity, as detected by the linker in this collection, help distinguish
relevant from non-relevant documents?} We call this \emph{Observable
Entity Relevance} (OER). OER is a property of the triple
(entity, query, collection): it measures whether entity $e$, as linked
in collection $\mathcal{D}$, is a discriminative signal for query $q$
in the BM25 candidate pool. Its complement, \emph{Conceptual Entity
Relevance} (CER), is the traditional notion of relevance: whether an
entity is semantically related to a query independent of the candidate
set or linking environment. CER underlies manually annotated entity
qrels and LLM-generated entity labels.

Binary-derived supervision (Equation~\ref{eq:binary_derivation}) sits
awkwardly with respect to both. It makes no semantic judgment, so it is
not CER; but its exclusivity criterion is a proxy for rarity rather than
for discriminative utility, so it is not OER either. As we show in
Section~\ref{sec:The Supervision--Coverage Dilemma}, it is in fact
\emph{negatively} correlated with OER. The three supervision families we
evaluate therefore target three different quantities --- exclusivity,
conceptual relatedness, and corpus-grounded discriminativeness --- and
none of them targets OER directly. We revisit the CER--OER distinction
in Section~\ref{sec:Observable vs Conceptual Entity Relevance}.

To make the distinction concrete, consider a query about \emph{gasoline
taxation policy}. \textsc{United States} is conceptually adjacent: it
appears in many relevant documents, but also in many non-relevant ones,
so its presence contributes little evidence of relevance. By contrast,
\textsc{Motor Fuel Tax} may be less globally salient, yet appear almost
exclusively in documents genuinely about fuel taxation. A CER-based
criterion would favor \textsc{United States}; OER reverses this ranking
because it rewards discriminative evidence in the collection rather than
semantic centrality alone.

We operationalize this using a log-odds difference, which measures how
much observing an entity shifts the likelihood that a document is
relevant. Entities that occur more often in relevant than non-relevant
documents receive positive scores, those that occur uniformly receive
scores near zero, and those that occur more often in non-relevant
documents receive negative scores. Because large differences can arise
from very few observations, we introduce a support term to discount
unreliable evidence. In this way, the formulation captures
discriminative evidence rather than raw prevalence.

\subsubsection{Formal Definition}
\label{subsubsec:oer_def}

For entity $e$, query $q$, and candidate pool
$\mathcal{D}_{\text{cand}}(q)$, we define:
\begin{equation}
\text{OER}(e, q) = w(e, q) \cdot \left[
\text{logit}\bigl(\hat{p}(e \mid \text{rel})\bigr) -
\text{logit}\bigl(\hat{p}(e \mid \text{nonrel})\bigr)
\right]
\label{eq:oer}
\end{equation}
The core term is the log-odds difference between entity presence rates
in relevant and non-relevant candidate documents: positive when the
entity appears more often in relevant documents, negative when it is an
anti-signal. The Laplace-smoothed probabilities are:
\begin{align}
\hat{p}(e \mid \text{rel}) &= \frac{df_{\text{rel}}(e, q) + \alpha}
{|\mathcal{R}_q \cap \mathcal{D}_{\text{cand}}(q)| + 2\alpha}
\label{eq:p_rel} \\
\hat{p}(e \mid \text{nonrel}) &= \frac{df_{\text{nonrel}}(e, q) +
\alpha}{|\overline{\mathcal{R}_q} \cap \mathcal{D}_{\text{cand}}(q)|
+ 2\alpha}
\label{eq:p_nonrel}
\end{align}
To discount extremely sparse evidence, we use the
support weight
\begin{equation}
w(e, q) = 1 - \exp\!\left(-\frac{df_{\text{cand}}(e, q)}{\tau}\right)
\label{eq:support}
\end{equation}
which down-weights entities that occur only a few
times in the candidate pool. OER is computed entirely from document-level relevance judgments and
entity linking outputs; it does not require manually annotated
entity-level relevance labels (\textit{entity qrels}).

A positive OER score indicates that $e$ is a discriminative signal; a
negative score indicates an anti-signal. OER is derived from the same
relevance judgments used for evaluation and is therefore not deployable:
it is never used as a training target, and no system we report as a
retrieval result uses it. Its role is diagnostic --- to measure whether a
given entity run selects observably discriminative entities, and to
quantify the gap between what supervision strategies select and what the
linked collection can support.

The one exception is the post-hoc filtering experiment of
Section~\ref{subsec:Attempting to Bridge the CER--OER Gap via OER Filtering}, where OER thresholds are applied to
an entity run at inference time. Those rows are an oracle upper bound on
what OER-aligned selection could achieve, not a deployable configuration,
and we label them as such.

\subsubsection{Signal Mode Taxonomy}
\label{subsubsec:Signal Mode Taxonomy}

To operationalize this diagnostic throughout the paper, we assign each
entity-query pair $(e,q)$ to one of five categories:
\begin{itemize}
    \item \textbf{Core signal}: high positive OER with adequate support
    ($df_{\text{cand}} > 2$, $df_{\text{rel}} \geq 2$,
    $df_{\text{rel}} > df_{\text{nonrel}}$, OER$(e,q) \geq 0.5$).

    \item \textbf{Conditional signal}: positive OER that does not meet
    core-signal criteria.

    \item \textbf{Generic bait}: OER$(e,q) \leq 0$ with high
    $df_{\text{cand}}$ ($\geq 50$); frequent but non-discriminative.

    \item \textbf{Anti-signal}: negative OER with
    $df_{\text{nonrel}} > df_{\text{rel}}$.

    \item \textbf{Sparse evidence}: $df_{\text{cand}} \leq 2$; too
    little candidate-pool support to be reliable.
\end{itemize}

We use \emph{bait rate} for the fraction of a run’s top-$k$ selections
classified as generic bait or anti-signal, \emph{signal rate} for the
fraction classified as core or conditional signal, and \emph{top-1 bait
rate} for the fraction of queries whose highest-ranked entity is bait.
These metrics connect the formal OER definition to the empirical
evidence in
Sections~\ref{sec:The Supervision--Coverage Dilemma}
and~\ref{sec:Observable vs Conceptual Entity Relevance}.

\subsection{Experimental Setup}
\label{subsec:Experimental Setup}

\subsubsection{Dataset}
We conduct all primary experiments on
TREC Robust04~\cite{voorhees2005robust}, an ad hoc retrieval benchmark
comprising 250 TREC topics over approximately 528K newswire
documents from the Financial Times, Federal Register, Foreign
Broadcast Information Service, and Los Angeles Times collections. All entity-oriented models in our evaluation use the
same entity-linked Robust04 collection, ensuring that performance
differences reflect model architecture and supervision strategy
rather than differences in the linking environment.

\subsubsection{Entity Linking and Entity Embeddings}
We annotate the Robust04 document collection using
WAT~\cite{piccinno2014wat}, a graph-based entity linker that
combines a coherence model over entity co-occurrences with a prior
probability model over entity surface forms. Entity
identifiers are Wikipedia page IDs. We use Wikipedia2Vec~\cite{yamada-etal-2020-wikipedia2vec} entity embeddings (300 dimensions) released by.

Linking density is a property of the evaluation environment that directly
bounds what the entity channel can reach, so we report it explicitly. Our
annotation of the 528{,}024-document collection yields 643{,}271 unique
entities and a mean of 116.0 entity mentions per document, corresponding
to 78.5 \emph{unique} entities per document (median 60). Over each query's
BM25 top-1000 candidate pool, the union contains a mean of 31{,}346 unique
entities (median 30{,}629; range 13{,}890--62{,}259); requiring an entity
to appear in at least two candidate documents reduces this to 14{,}119.
All coverage and discrimination statistics in this paper are computed over
the unfiltered top-1000 pool and count distinct entities.\footnote{The
QDER paper~\cite{chatterjee2025qder} reports 116 entities per document for
this collection; that figure counts mentions. We report unique entities
here, since coverage depends on distinct entities rather than mention
frequency.} Linker configuration materially changes these quantities, so
effectiveness figures for entity-oriented methods are comparable across
studies only when the linking environment is held fixed or reported.

\subsubsection{First-stage Retrieval}
All experiments use Lucene BM25 (default parameters) with RM3 query expansion as the first-stage retriever, producing a candidate pool of the top-1000 documents per query. All coverage statistics, OER scores, and entity run evaluations are computed over this BM25 top-1000 candidate pool. 

\subsubsection{Neural Entity-Oriented Document Re-Ranking Models}

We evaluate six neural entity-oriented architectures alongside a BM25 baseline \emph{as re-rankers}:

\begin{itemize}

\item \textbf{EDRM}~\cite{liu-etal-2018-entity}: Entity-Duet
Neural Ranking Model. Integrates entity embeddings from a
knowledge graph into neural document ranking via a dual
word-entity interaction structure with kernel pooling.

\item \textbf{Word-Entity Duet}~\cite{xiong2017word}: Models
four-way interactions between query and document terms and
entities through dual attention-based representation learning.

\item \textbf{EVA}~\cite{tran2022dense}: Dense retrieval with
entity views. Clusters document entities into semantically
coherent views using Wikipedia2Vec similarity and enriches
dense BERT-based query-document representations with
entity-view embeddings indexed for approximate nearest
neighbour search.

\item \textbf{EsdRank}~\cite{xiong2015esdrank}: Connects
queries and documents through external semi-structured
knowledge base data, constructing entity-enriched document
representations via knowledge graph traversal.

\item \textbf{DREQ}~\cite{chatterjee2024dreq}: Document
re-ranking using entity-based query understanding. Uses an
explicit MonoBERT entity ranking stage to construct
query-specific entity representations that guide
cross-encoder document scoring.

\item \textbf{QDER}~\cite{chatterjee2025qder}: Query-specific
document and entity representations. Uses a MonoBERT entity
ranker to produce query-specific entity scores, which are
used to construct multi-vector document representations for
re-ranking via a learned interaction model.
\end{itemize}

For EsdRank, we could not find a public codebase and hence use our own reimplementation based on the descriptions provided in the
original papers. For other models, we use the officialy released code. 

\subsubsection{Unsupervised Entity-Oriented Document Re-Ranking Models}
\label{subsubsec:unsupervised_doc_runs}

To complement supervised models, we construct a broad suite of
\emph{unsupervised} entity-based document ranking methods that rerank a
common BM25 candidate pool using query-specific entity signals. These
methods take as input a BM25 run and an entity ranking, and incorporate
entity information without any learned parameters.

We construct a diverse set of methods to explore how entity signals can
be incorporated into document ranking. We consider several classes of
approaches: (i) direct integration methods that combine entity scores
with BM25 or use entity overlap, (ii) query expansion methods that use
entity descriptions to augment the query, (iii) representation-based
methods that rank using entity embeddings (e.g., Wikipedia2Vec), and
(iv) probabilistic and graph-based methods that propagate entity signals
through co-occurrence or similarity structures.

For each method, we perform a systematic sweep over key parameters such
as the number of query entities ($k$), interpolation weights, and
propagation settings, yielding \textbf{437 distinct runs} evaluated under
a unified pipeline. These are \emph{document-ranking} configurations and
are counted separately from the entity-selection methods of
Section~\ref{subsubsec:entity_rankers}; the two sets are disjoint, and
figures reported for one do not include the other.
This sweep allows us to test whether the limitations identified earlier
persist across diverse ways of using entity signals. As we show later,
the same patterns emerge consistently, indicating that the limitations
are structural rather than model-specific. Full details of all methods
and configurations are provided in the online appendix.

\subsubsection{Entity Rankers and Supervision Strategies}
\label{subsubsec:entity_rankers}

To analyze supervision noise (Section~\ref{sec:The Supervision--Coverage Dilemma}), we evaluate five entity rankers forming a ladder of increasing supervision quality. Each ranker is applied to an unfiltered BM25 top-1000 entity pool, and the resulting top-20 entities are used for document ranking under both conditional and open-world settings. The document ranker is fixed across configurations; only the entity supervision and selection strategy varies.

\textbf{(1) BERT (binary-derived, qrels).}
The strongest supervised configuration available under standard
benchmarks. We use the label \emph{binary-derived (qrels)} rather than
\emph{oracle}, since the supervision is derived from the document
relevance judgments themselves rather than from independent entity-level
annotation.  Entity
qrels are derived from document relevance judgments using the binary rule
in Equation~\ref{eq:binary_derivation}, retaining exclusive positives
$\mathcal{E}^+_q$ and exclusive negatives $\mathcal{E}^-_q$ and discarding
common entities $\mathcal{E}^0_q$. A MonoBERT cross-encoder entity ranker
is trained on these derived qrels using five-fold cross-validation over
the 250 Robust04 topics, with entity qrels derived separately per fold to
prevent leakage and negatives balanced to match the number of positives.
At test time, the frozen ranker is applied to the full unfiltered BM25
top-1000 entity pool and the top-20 entities are selected for document
ranking. Documents sharing zero entities with this top-20 run are removed
for conditional evaluation; all BM25 top-1000 documents are retained for
open-world evaluation.


\textbf{(2) BERT (LLM CER).}
A MonoBERT entity ranker trained on entity relevance labels generated by
three LLM judges: Claude Haiku 4.5, GPT-4o Mini, and Gemma 3 27B.
Query--entity pairs for annotation were drawn from the top-100 entities per query
from an unsupervised entity ranking method. The ranker
combines corpus-IDF-weighted SBERT entity--entity similarity with SBERT
query--entity text similarity, with both scores normalised before
combination.

For each query--entity pair, the LLM is given the query title,
description, and narrative together with the entity's Wikipedia title and
first-paragraph description, and asked to judge whether the entity is a
useful \emph{topical signal} for the query on a three-point scale: 2 for
core topical matches, 1 for semantically related supporting signals, and 0
for generic, irrelevant, or anti-signal entities. The prompt explicitly
instructs the LLM to avoid rewarding generic concepts. Because the LLM
has no access to candidate documents, corpus statistics, or entity linking
behaviour, these labels capture \emph{Conceptual Entity Relevance} (CER)
rather than observable discriminativeness.

Label stability across the three judges: pairwise Cohen's $\kappa$ of
0.559 (Haiku vs.\ GPT-4o Mini), 0.411 (Haiku vs.\ Gemma), and 0.431
(GPT-4o Mini vs.\ Gemma); Fleiss' $\kappa = 0.445$; full three-way
agreement on 74.5\% of examples. Majority-vote labels are used as CER
supervision for the MonoBERT ranker.

\textbf{(3) BERT (LLM CER + IDF).}
The LLM CER ranker with post-hoc in-set IDF adjustment applied at
inference time. For each query, a local IDF weight is computed from the
BM25 top-1000 candidate pool:
\begin{equation}
\text{IDF}_{\text{local}}(e, q) =
\log\frac{N_q + 1}{df_{\text{cand}}(e, q) + 1} + 1
\label{eq:local_idf}
\end{equation}
Entity scores are rescaled as $\text{score}'(e, q) = \text{score}(e, q)
\times \text{IDF}_{\text{local}}(e, q)$, with entities absent from the
candidate pool receiving a score of zero. This down-weights
high-frequency entities that appear broadly across the candidate pool,
reducing generic bait in the top-20 selection without modifying the
trained model.

\textbf{(4) LGBM entity selector.}
A LightGBM learning-to-rank model that reranks the top-100 candidate
entities from the first stage to produce the top-20 selection. Features fall into
three groups: (i) \emph{candidate support}---raw and deduplicated
document frequency in the BM25 candidate pool, total mention count, and
rank-weighted support ($\sum_d 1/\text{rank}_d$); (ii) \emph{stage-1
prior}---first-stage entity score and rank; and (iii) \emph{lexical
overlap}---token overlap count, Jaccard coefficient, and query coverage.
All features are computable at test time without relevance judgments.
Qrels-derived features are explicitly excluded to prevent supervision
leakage---these include $df_{\text{rel}}$, $df_{\text{nonrel}}$,
precision, lift, log-odds, relevance coverage, and all OER-derived scores.
The model is trained with a LambdaRank objective using 5-fold CV.

\textbf{(5) Consensus entity ranker.}
An unsupervised method that operates entirely from the
BM25 candidate pool without any relevance labels. A consensus gate first
removes entities appearing in fewer than two candidate documents, on the
grounds that linker errors are idiosyncratic and rarely corroborated
across documents. Surviving entities are scored by:
\begin{equation}
\text{score}(e, q) = \text{soft\_support}(e, q) \times
\left( \log\frac{K+1}{df_{\text{cand}}(e,q)+1} + 1 \right)
\label{eq:consensus}
\end{equation}
where the specificity term is a pseudo-IDF over the candidate neighbourhood
--- entities concentrated in a few top-ranked documents are upweighted
over entities appearing in all $K$ documents. The soft support term
weights each document's contribution to entity $e$'s score. We evaluate
three variants: \textit{Consensus (rho)}, which weights each document by
the maximum WAT linker confidence $\rho$ assigned to entity $e$ in that
document; \textit{Consensus (rank)}, which weights by reciprocal BM25
log-rank ($1/\log_2(\text{rank}_d + 1)$); and \textit{Consensus
(rho+rank)}, which multiplies both weights.

\subsubsection{Model-Specific Evaluation Protocols}
\label{subsubsec:model_eval_protocols}

For QDER and DREQ, the entity-run and document-pool choices are varied
independently, since they are separable and differ by two orders of
magnitude in effect. Entity qrels are first derived
from document qrels using the binary rule in
Equation~\ref{eq:binary_derivation}, split query-wise into five folds,
and used to train a BERT entity ranker. For the conditional setting, the
entity ranker is trained and evaluated by cross-validation on these
fold-specific data, and the resulting fold-specific top-$20$ entity runs
are used to construct document-ranking data by removing documents that
share no entity with the top-$20$ selection. The document ranker is then
trained once on this conditional training data.

For the open-world setting, no retraining is performed. The same frozen
entity ranker is applied to an unfiltered entity pool induced by all
entities appearing in the BM25 top-1000 documents for each query, and
the resulting top-$20$ entity run is used to construct open-world test
data without removing zero-overlap documents. The document ranker is
also kept fixed: it is trained once on the conditional training data and
then evaluated on both the conditional and open-world test sets. This
holds the document model constant and isolates the effect of changing the
entity run and evaluation pool.

For entity-aware models without an explicit entity-ranking stage (EDRM, EVA, etc.), the distinction is defined at the document
candidate-set level rather than through a separate entity run. In the
conditional setting, evaluation is restricted to BM25 candidates with at
least one usable linked entity. In the open-world setting, the full BM25
candidate pool is retained, including documents with empty, sparse, or
noisy entity evidence. The architecture is unchanged; only the
evaluation pool differs. This provides the closest analogue of the
conditional/open-world distinction for models whose entity channel is
integrated directly into the document ranker.

\subsubsection{Additional Unsupervised Entity Ranking Configurations}
\label{subsubsec:unsupervised_entity_runs}

In addition to the consensus ranker in Section~\ref{subsubsec:entity_rankers}, we construct a broad suite of
\emph{unsupervised} entity ranking configurations to explore how entity
signals can be selected without learned supervision. Each method
produces a query-specific ranked list of entities from a shared BM25
candidate pool, which is truncated to the top-$k$ ($k=20$) for downstream
document ranking and analysis.

The methods span three main approaches. \textbf{Similarity-based methods}
rank entities using embedding or text similarity to the query or its
linked entities, with variants incorporating IDF weighting or coverage
signals. \textbf{Graph-based methods} propagate relevance over a local
entity graph (e.g., Personalized PageRank), capturing multi-hop
relationships. \textbf{LLM-based methods} re-rank entities using
instruction-tuned models based on query text and entity descriptions.
Hybrid methods combine entity-space and text-space similarity signals.

We perform a systematic sweep over key design choices, including
embedding source, similarity function, aggregation strategy, and score
combination. This yields \textit{187 distinct runs} over the same
candidate pool. These runs allow us to test whether the limitations observed later in
the paper persist across diverse ways of selecting entity signals, and
to separate the quality of the entity signal from the behavior of the
downstream ranking model. Detailed method descriptions with full configuration
details in the online appendix.

\subsubsection{Evaluation Metrics}
For document ranking we report MAP, nDCG@20, and P@20 using
\texttt{trec\_eval} against Robust04 relevance judgments, treating
grades 1 and 2 as relevant. For coverage analysis we report
$\text{RelCov}@k$, $\text{NonRelCov}@k$, and
$\text{DiscRatio}@k$ for $k \in \{10, 20, 50\}$, computed over
the BM25 top-1000 candidate pool. For supervision noise analysis
we report bait rate, signal rate, and top-1 bait rate as defined
in Section~\ref{subsec:Observable Entity Relevance}. All entity ranker experiments use
five-fold cross-validation over the Robust04 title query set to avoid
train/test leakage in entity qrel derivation, with model selection
on the validation fold and metrics reported on the held-out test
fold.

\section{The Entity Run, Not the Document Pool, Determines Effectiveness}
\label{sec:Open-World Evaluation Reveals Collapse}

Entity-oriented ranking is motivated by the claim that query-relevant
entities can provide useful evidence for document retrieval. We test this
claim directly on Robust04 across six entity-aware architectures,
separating two choices that prior work conflates: whether the entity run
is derived from the relevance judgments used for evaluation, and whether
the document pool is restricted to entity-matched candidates. The result
is an immediate inconsistency: the same models appear highly effective
under one supervision regime and largely ineffective under the other,
while the document pool proves almost irrelevant. The remainder of this section
establishes what drives that inconsistency. It is not the ranking
architecture. It originates on the entity side, depends on how the
entity signal is constructed, and persists across a large design space.

\subsection{Main Results}
\label{subsec:main_results}

Table~\ref{tab:main_results} reports MAP and nDCG@20 under both
supervision regimes. Under leaked entity supervision, QDER reaches MAP
0.608 and DREQ reaches 0.697, far above BM25 at 0.292 and consistent
with the figures reported in the original papers. Under clean entity
supervision, both collapse to BM25 level: QDER drops to 0.294 and DREQ
to 0.293. For QDER and DREQ the document pool is the full BM25 top-1000
in both columns, with unrestricted recall denominators; their collapse is
not produced by changing what is ranked. The remaining models have no
entity-ranking stage and never use qrels-derived entity labels, so for
them the two columns differ only in whether the document pool is
restricted. EDRM, Word-Entity Duet, EVA, and EsdRank are weak in both
columns and remain well below BM25---and the difference between the
columns is negligible in every case, which is independent evidence that
the document pool is not what drives effectiveness.

The key finding is not merely that open-world numbers are lower. It is
that the two settings yield sharply different conclusions about the same
models. Under leaked entity supervision, entity-oriented methods appear to
produce dramatic improvements. Under open-world evaluation, those gains
largely disappear. As argued in Section~\ref{sec:Introduction}, these regimes answer
different questions. Leaked supervision asks whether a model can exploit
entity evidence \emph{when that evidence is already aligned with
relevance}. Clean supervision asks whether the pipeline improves ranking
when that alignment must be inferred from the collection alone. The two
questions are not equivalent, and the gap between their answers is the
central phenomenon this paper diagnoses.

\begin{table}[t]
\centering
\caption{Effectiveness on Robust04 under two distinct interventions. For
models without an entity-ranking stage (upper panel), the columns differ
in whether the document pool is restricted to candidates containing a
usable linked entity. For models with an entity ranker trained on
qrels-derived labels (lower panel), the columns differ in whether that
entity run is derived from the evaluation judgments; \textbf{the document
pool is the full BM25 top-1000 in both columns of the lower panel}. All
figures use \texttt{trec\_eval -c} against the full judgment set. Best
result under clean supervision in \textbf{bold}.}
\label{tab:main_results}
\begin{tabular}{lcccc}
\toprule
& \multicolumn{2}{c}{\textbf{Restricted / Leaked}} &
  \multicolumn{2}{c}{\textbf{Full / Clean}} \\
\cmidrule(lr){2-3}\cmidrule(lr){4-5}
\textbf{Model} & MAP & nDCG@20 & MAP & nDCG@20 \\
\midrule
BM25+RM3         & ---   & ---   & 0.292 & 0.435 \\
\midrule
\multicolumn{5}{l}{\textit{No entity-ranking stage---document pool varies}} \\
EDRM-KNRM        & 0.089 & 0.153 & 0.089 & 0.151 \\
EDRM-ConvKNRM    & 0.087 & 0.150 & 0.088 & 0.150 \\
Word-Entity Duet & 0.151 & 0.241 & 0.149 & 0.235 \\
EVA              & 0.156 & 0.312 & 0.167 & 0.302 \\
EsdRank          & 0.124 & 0.213 & 0.111 & 0.202 \\
\midrule
\multicolumn{5}{l}{\textit{Qrels-derived entity ranker---entity run varies, pool held full}} \\
DREQ             & 0.697 & 0.867 & 0.293 & 0.439 \\
QDER             & 0.608 & 0.769 & 0.294 & 0.438 \\
\bottomrule
\end{tabular}
\end{table}

\subsection{The Architecture Is Not the Explanation}
\label{subsec:architecture_not_problem}

Before diagnosing the inconsistency, it is important to establish that
it does not arise from the ranking architecture being weak.
Table~\ref{tab:neural_comparison} compares the best open-world
entity-oriented result against a range of neural rerankers on the same
collection. The best open-world configuration---QDER with the LGBM
entity selector---achieves MAP 0.34, matching the official TREC
Robust04 best system (MAP 0.33) and outperforming the majority of neural baselines,
including ColBERT v2, RankVicuna, and most fine-tuned
cross-encoders. The architecture is competitive when evaluated on equal
terms.

The same QDER architecture reports MAP 0.60 in its original paper. That
figure is arithmetically correct and was computed over the full 1000-document
candidate pool with unrestricted recall denominators---it is not the
product of a restricted evaluation set. What it measures is model
capacity when the entity run is aligned with the evaluation judgments,
which is a different quantity from MAP 0.34 under clean supervision. A
system can produce 0.60 under leaked supervision and 0.29 under clean
supervision simultaneously---as the binary-derived QDER run does---because the entities the judgments identify are highly discriminative but
cover only a small fraction of relevant documents. Conversely, a system
can reach 0.34 under clean supervision---as the QDER (LGBM) run does---because its entities are broad enough to be selectable without
judgments at all. The gap between the two is not a measure of model
failure. It is a measure of how much of the entity signal was supplied by
the judgments rather than recovered from the collection.

\begin{table}[t]
\centering
\caption{Open-world MAP on Robust04 for a range of neural reranking
models, compared against the best QDER open-world configuration.
$\blacktriangle$ / $\blacktriangledown$ denote statistically significant
improvement / degradation over BM25.}
\label{tab:neural_comparison}
\begin{tabular}{llccc}
\toprule
\textbf{Model} & \textbf{Type} & \textbf{MAP} & \textbf{nDCG@20} &
\textbf{P@20} \\
\midrule
BM25 & Lexical & 0.292 & 0.435 & 0.384 \\
\midrule
\multicolumn{5}{l}{\textit{Cross-Encoder Models (Fine-Tuned)}} \\
RankT5~\cite{zhuang2023rankt5}      & Cross-encoder & 0.303 & 0.494$\blacktriangle$ & 0.429$\blacktriangle$ \\
MonoBERT~\cite{nogueira2019passage} & Cross-encoder & 0.297 & 0.479 & 0.409 \\
RoBERTa~\cite{liu2019roberta}       & Cross-encoder & 0.290 & 0.474 & 0.410 \\
DeBERTa~\cite{he2020deberta}        & Cross-encoder & 0.293 & 0.486 & 0.422 \\
ELECTRA~\cite{clark2020electra}     & Cross-encoder & 0.268 & 0.446 & 0.387 \\
ERNIE~\cite{zhang-etal-2019-ernie}  & Cross-encoder & 0.289 & 0.475 & 0.412 \\
\midrule
\multicolumn{5}{l}{\textit{Bi-Encoder Models}} \\
ColBERT v2~\cite{khattab2020colbert} & Bi-encoder (FT) & 0.292 & 0.473$\blacktriangle$ & 0.410$\blacktriangle$ \\
DPR~\cite{karpukhin-etal-2020-dense} & Bi-encoder (ZS) & 0.170$\blacktriangledown$ & 0.300$\blacktriangledown$ & 0.259$\blacktriangledown$ \\
\midrule
\multicolumn{5}{l}{\textit{LLM-Based Models (Zero-Shot)}} \\
RankVicuna~\cite{pradeep2023rankvicuna} & LLM listwise & 0.296 & 0.467 & 0.400 \\
RankZypher~\cite{pradeep2023rankzephyr} & LLM listwise & 0.318$\blacktriangle$ & 0.505$\blacktriangle$ & 0.433$\blacktriangle$ \\
\midrule
\multicolumn{5}{l}{\textit{Entity-Oriented (this work)}} \\
Best open-world (QDER) & Entity-aware & \textbf{0.343} & \textbf{0.491} & \textbf{0.430} \\
\bottomrule
\end{tabular}
\end{table}

\subsection{The Inconsistency Depends on the Entity Signal}
\label{subsec:supervision_dependent}

If the inconsistency is not architectural, the natural question is
whether it depends on how the entity signal is constructed.
Table~\ref{tab:supervision_ladder} answers this by fixing the document
ranking architecture at QDER and varying only the entity run source.

The binary-derived entity run collapses from MAP 0.608 under leaked
supervision to 0.294 under clean supervision---a gap of 0.314 points,
larger than BM25's absolute MAP. The LLM CER, LLM CER+IDF, Consensus, and
LGBM runs are constructed without relevance judgments and so admit no
leaked variant; for these, the two columns differ only in whether the
document pool is restricted, and the difference is at most 0.010 MAP. The
absence of collapse is not the same as strong effectiveness: these runs
reach MAP between 0.306 and 0.343, only modestly above BM25.

This reveals the underlying tension directly. Entity runs that the
judgments identify are narrow and highly discriminative, but cover too
few relevant documents to help once those judgments are withdrawn. Entity
runs recoverable without judgments are broad enough to cover most of the
candidate pool, but that same breadth dilutes discriminative power. The
inconsistency is therefore a property of the entity signal, not of the
architecture, and the two regimes sit at opposite ends of the same
coverage--discrimination tradeoff.

\begin{table}[t]
\centering
\caption{QDER under different entity supervision strategies; architecture
and document ranker held fixed. For the binary-derived run the two
columns differ in the entity run (derived from the evaluation judgments
vs.\ constructed without them). The remaining runs use no relevance
judgments and admit no leaked variant, so for those rows the columns
differ only in whether the document pool is restricted to entity-matched
candidates --- a difference of at most 0.010 MAP. All figures use
\texttt{trec\_eval -c} against the full judgment set.}
\label{tab:supervision_ladder}
\begin{tabular}{lcccc}
\toprule
& \multicolumn{2}{c}{\textbf{Restricted / Leaked}} &
  \multicolumn{2}{c}{\textbf{Full / Clean}} \\
\cmidrule(lr){2-3}\cmidrule(lr){4-5}
\textbf{Entity run source} & MAP & nDCG@20 & MAP & nDCG@20 \\
\midrule
BM25+RM3 (baseline)   & ---   & ---   & 0.292 & 0.435 \\
\midrule
BERT (binary-derived) & 0.608 & 0.769 & 0.294 & 0.438 \\
BERT (LLM CER)        & 0.310 & 0.438 & 0.310 & 0.464 \\
BERT (LLM CER+IDF)    & 0.310 & 0.457 & 0.308 & 0.456 \\
Consensus (rho+rank)  & 0.306 & 0.457 & 0.306 & 0.455 \\
LGBM Selector         & 0.333 & 0.461 & 0.343 & 0.491 \\
\bottomrule
\end{tabular}
\end{table}

\subsection{The Bottleneck Is Entity-Side, Not Document-Side}
\label{subsec:entity_side_bottleneck}

One alternative explanation is that open-world evaluation simply makes
the ranking problem harder by expanding the document candidate pool.
Table~\ref{tab:dreq_filtering} rules this out using DREQ, which allows
the entity side and document side to be varied independently.

We observe that holding the entity run leaked, MAP
stays at 0.695--0.698 regardless of whether the document pool is
filtered to entity-matched documents or expanded to the full BM25
top-1000. Moving to a clean entity run, MAP collapses to
0.291--0.293 regardless of document-side filtering. The document-side
decision contributes at most 0.003 MAP points; the entity-side decision
contributes 0.405. Recall@1000 is identical across all four conditions
(0.7772), confirming that the underlying candidate pool is unchanged and
that no relevant documents are being lost. The collapse is caused
entirely by the degradation of entity signal quality once selection must
operate over the full open-world entity distribution.

\begin{table}[t]
\centering
\caption{DREQ performance under four combinations of entity-side and
document-side evaluation. Recall@1000 is identical across all
conditions, confirming the candidate pool is unchanged.}
\label{tab:dreq_filtering}
\begin{tabular}{llcccc}
\toprule
\textbf{Entity side} & \textbf{Doc side} & \textbf{MAP} &
\textbf{nDCG@10} & \textbf{nDCG@20} & \textbf{Recall@1000} \\
\midrule
Leaked & Closed-world & 0.6947 & 0.8819 & 0.8675 & 0.7772 \\
Leaked & Open-world   & 0.6975 & 0.8802 & 0.8670 & 0.7772 \\
Clean   & Closed-world & 0.2911 & 0.4503 & 0.4348 & 0.7772 \\
Clean   & Open-world   & 0.2927 & 0.4556 & 0.4390 & 0.7772 \\
\bottomrule
\end{tabular}
\end{table}

\subsection{Filtering and Balancing Move Scores in Opposite Directions}
\label{subsec:filtering_vs_balancing}

Because document-side interventions are frequently assumed to inflate
effectiveness, it is worth stating that under \texttt{trec\_eval -c}
against the full judgment set they do not. A run restricted to
entity-matched candidates is charged for every relevant document it
discards. On Robust04 (title), the QDER neural stage restricted in this
way scores MAP 0.5667, against 0.6082 for the same model over the full
1000-document pool: restriction costs 0.042 MAP.

Class balancing behaves oppositely. Balancing retains all positive
candidates and downsamples negatives, so the recall denominator is
largely preserved while most of the ranking task is removed. Applied to
evaluation candidates it inflates rather than deflates.

The two operations are therefore not interchangeable, and reports that do
not state which was applied---and under which \texttt{trec\_eval}
flag---are not interpretable. This distinction matters for interpreting
the pipeline: the entity-derived candidate filter used by QDER and DREQ
retains 93.8\% of relevant candidates but only 44.6\% of non-relevant
ones, since the entity set is derived from the document judgments. The
resulting pool is enriched for relevant documents, yet scoring it under
\texttt{-c} still costs MAP relative to ranking everything. What raises
effectiveness is the entity run, not the pool.

\subsection{The Pattern Holds Across 437 Configurations}
\label{subsec:437_configs}

The preceding results involve a small number of architectures and
supervision strategies. To test whether the ceiling is a broader
phenomenon, we evaluate 437 unsupervised entity-oriented document
ranking configurations spanning a wide range of entity selection
methods, weighting schemes, and scoring functions. These are
\emph{document-ranking} configurations and are disjoint from the
entity-selection methods of Table~\ref{tab:supervision_ladder}; the
Consensus entity ranker reported there feeds the QDER document ranker and
is not among the 437. Not a single one of the 437 surpasses BM25 under
clean supervision. All fall below MAP 0.292, with mean MAP 0.231, median
0.241, and maximum 0.289. The supervised configurations in Table~\ref{tab:main_results} do not
break this ceiling either: even the best supervised open-world result
improves MAP by only 0.051 over BM25.

The inconsistency identified in this section is therefore not a
peculiarity of any particular model, supervision strategy, or entity
selection method. It is consistent across a large design space. The
ceiling on open-world performance is shared, and it is not resolved by
changing the architecture or the supervision. The next section explains why this ceiling arises, tracing it to the
supervision--coverage dilemma in the entity channel.

\section{The Supervision--Coverage Dilemma}
\label{sec:The Supervision--Coverage Dilemma}

Section~\ref{sec:Open-World Evaluation Reveals Collapse} established that
entity-oriented ranking fails uniformly under clean entity supervision and
that the bottleneck is on the entity side. We now explain why. The failure arises from two structural constraints built into current
benchmark construction. The
first is a coverage ceiling: entity linking is incomplete, so many
relevant documents do not expose the entities that would be needed to
support them through the entity channel. The second is a supervision
problem: standard benchmarks provide no entity-level relevance
annotations, so entity rankers must be trained on proxy labels derived
from document relevance --- labels that are noisy at the source and
further corrupted during training. Together, these two constraints create
a structural dilemma: the entity selections that are clean enough to
discriminate are too sparse to cover most relevant documents, while the
selections broad enough to cover relevant documents are too noisy to
discriminate. No configuration escapes this tradeoff.

\subsection{The Coverage Ceiling}
\label{subsec:The Coverage Ceiling}

For entity signals to improve document ranking, they must first reach
the documents being ranked. Table~\ref{tab:coverage} shows that no
evaluated configuration achieves both high relevant-document coverage
and strong discrimination simultaneously.

Selecting entities directly by their measured discriminativeness on the
test judgments---information no deployable system has---reaches
RelCov@20 $= 72.8\%$ at DiscRatio $= 5.40$, and $87.1\%$ at DiscRatio
$= 3.37$ for $k = 50$. This is the only configuration that approaches the
desirable region, and it does so using precisely the signal the benchmark
fails to annotate.

The binary-derived run sits at the opposite extreme: DiscRatio@20 $= 154.9$ but
RelCov@20 $= 19.7\%$, leaving 80.3\% of relevant documents unreachable
through the entity channel regardless of downstream ranker quality. All
other runs occupy the opposite regime. LLM CER covers 97.1\% of relevant
documents but achieves DiscRatio@20 $= 1.09$ --- its top-20 entities
appear in relevant and non-relevant documents at nearly identical rates.
IDF filtering shifts this only marginally (DiscRatio 1.23, RelCov
91.2\%). The LGBM selector gives the best leakage-free compromise (RelCov
94.9\%, DiscRatio 1.20), while consensus variants cluster at similarly
high coverage and near-random discrimination.

\begin{table}[t]
\centering
\caption{Coverage and discrimination at $k=20$ over the BM25 top-1000
candidate pool. RelCov and NonRelCov are computed exactly at document
level, with the denominator restricted to relevant (resp.\ non-relevant)
documents within the candidate pool. The two oracle rows use the test
relevance judgments and are not deployable; they bound what entity
selection could achieve if observable discriminativeness were annotated.
No leakage-free configuration achieves both high relevant-document
coverage and high discriminative precision. The binary-derived row
establishes the ceiling under the strongest supervision these benchmarks
can produce without external annotation: 19.7\% of relevant documents. }
\label{tab:coverage}
\begin{tabular}{lcccc}
\toprule
\textbf{Run} & \textbf{RelCov@20} & \textbf{NonRelCov@20} &
\textbf{DiscRatio@20} & \textbf{Mean overlap} \\
\midrule
\multicolumn{5}{l}{\textit{Oracle diagnostics --- use test judgments, not deployable}} \\
OER log-odds selection         & 0.728 & 0.135 & 5.40  & 3.09 \\
BERT (binary-derived, qrels)   & 0.197 & 0.001 & 154.9 & 1.69 \\
\midrule
\multicolumn{5}{l}{\textit{Leakage-free selection}} \\
BERT (LLM CER)                 & 0.971 & 0.888 & 1.09  & 4.33 \\
BERT (LLM CER + IDF)           & 0.912 & 0.741 & 1.23  & 3.33 \\
LGBM selector                  & 0.949 & 0.787 & 1.20  & 4.24 \\
Consensus (rho+rank)           & 0.959 & 0.908 & 1.06  & 5.48 \\
\bottomrule
\end{tabular}
\end{table}

The tradeoff is not resolved by selecting more entities. As $k$ increases
from 10 to 50, RelCov rises for every run but DiscRatio falls in
parallel. For LGBM, RelCov rises from 85.6\% to 99.6\% while DiscRatio
falls from 1.44 to 1.03; for the binary-derived run, from 11.8\% to
33.8\% while DiscRatio falls from 211.9 to 96.4. This is a structural
slope present across all values of $k$ and all runs, not a tuning
problem. Mean overlap reinforces this: even when the binary-derived run
reaches a relevant document, it contributes on average only 1.69
supporting entities, compared with 4.24 for LGBM and 4.33 for LLM CER.
The signal is sparse not only across documents but within them.

\begin{figure}[t]
    \centering
    \includegraphics[width=\linewidth]{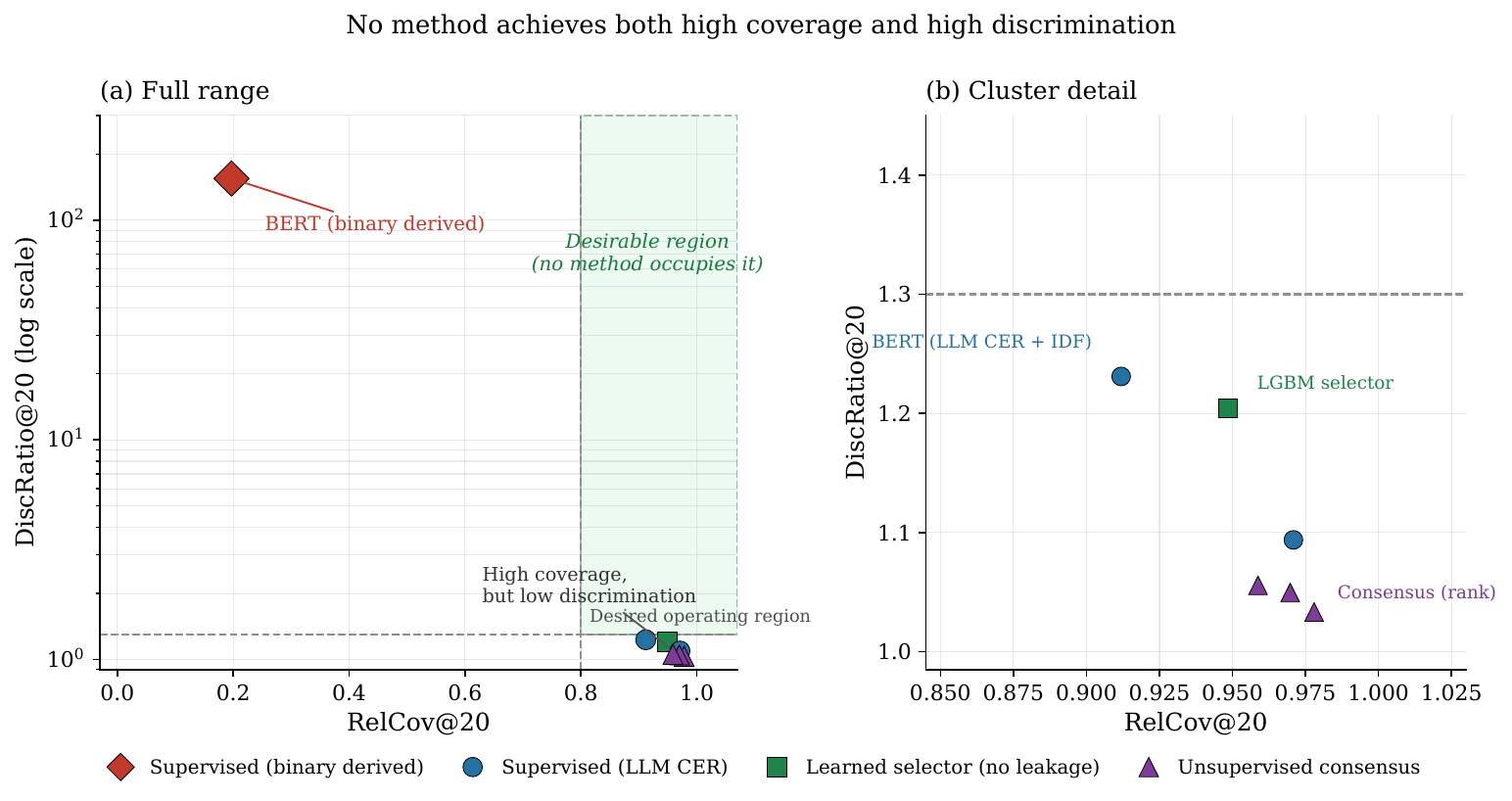}
    \caption{Coverage--discrimination space for the evaluated entity
    selection configurations at $k=20$. Panel~(a): full range on a
    log-scale DiscRatio axis. The binary-derived run achieves near-perfect
    discrimination (DiscRatio $= 154.9$) but covers only 19.7\% of relevant
    documents, while all other configurations cluster in the lower-right with
    high coverage and low discrimination. Panel~(b): zoom into the
    high-coverage cluster. Even the best leakage-free configurations fall below
    DiscRatio $= 1.3$, and the desirable region (upper-right, shaded) remains
    empty across all configurations.}
    \label{fig:coverage_discrimination}
\end{figure}

\begin{figure}[t]
    \centering
    \includegraphics[width=\linewidth]{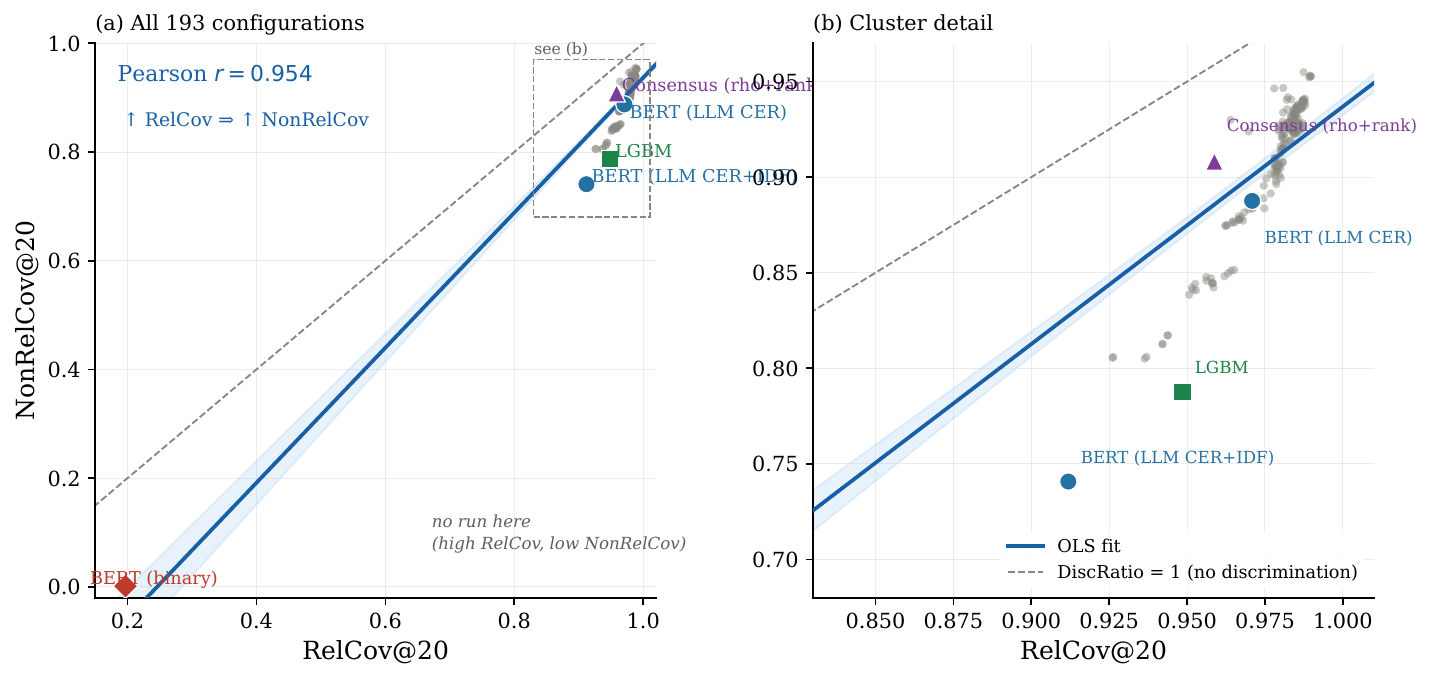}
    \caption{RelCov@20 vs.\ NonRelCov@20 across all 193 entity selection
    configurations ($r = 0.954$, $p < 0.001$). Panel~(a): full range showing
    the strong linear trend across all method families. Panel~(b): detail of
    the dense cluster, with the five representative runs from
    Table~\ref{tab:coverage} marked. No configuration occupies the upper-left
    region of high RelCov with low NonRelCov.}
    \label{fig:relcov_nonrelcov}
\end{figure}

Figure~\ref{fig:coverage_discrimination} summarises the picture.
Figure~\ref{fig:relcov_nonrelcov} extends it to all 193 configurations:
RelCov@20 and NonRelCov@20 are near-perfectly correlated across the full
space ($r = 0.954$, $p < 0.001$, slope $= 1.24$). Every gain in
relevant-document coverage is accompanied by a proportional rise in
non-relevant coverage, regardless of method family or tuning. No
configuration occupies the desirable upper-left region of high RelCov
with low NonRelCov. The coverage ceiling is a property of the Robust04
entity linking environment, not of any particular model.

\subsection{Supervision Noise}
\label{subsec:Supervision Noise}

The coverage ceiling explains one side of the dilemma. The other is that
even within the reachable portion of the candidate pool, entity signals
are often non-discriminative. Table~\ref{tab:bait} shows bait rates
across the supervision ladder.

\begin{table}[t]
\centering
\small
\caption{Bait rate, signal rate, and top-1 bait rate across the
supervision ladder at $k=20$. Binary-derived supervision achieves near-zero bait
but only at the cost of covering 19.7\% of relevant documents. Better
supervision reduces bait but does not eliminate it, and does not escape
the coverage--discrimination tradeoff. RelCov@20 is computed exactly at document level over the BM25 top-1000
candidate pool, as in Table~\ref{tab:coverage}.}
\label{tab:bait}
\begin{tabular}{lcccc}
\toprule
\textbf{Run} & \textbf{Bait rate} & \textbf{Signal rate} &
\textbf{Top-1 bait rate} & \textbf{RelCov@20} \\
\midrule
BERT (LLM CER)        & 0.292 & 0.667 & 0.096 & 0.971 \\
BERT (LLM CER + IDF)  & 0.312 & 0.587 & 0.185 & 0.912 \\
Consensus (rank)      & 0.245 & 0.725 & 0.240 & 0.978 \\
Consensus (rho)       & 0.255 & 0.707 & 0.240 & 0.970 \\
Consensus (rho+rank)  & 0.213 & 0.726 & 0.196 & 0.959 \\
LGBM selector         & 0.150 & 0.786 & 0.080 & 0.949 \\
\midrule
BERT (binary-derived, qrels) & 0.002 & 0.074 & 0.000 & 0.197 \\
\bottomrule
\end{tabular}
\end{table}

Three observations follow. First, bait is persistent across the full
supervision ladder. The best leakage-free configuration---LGBM, trained
on corpus-grounded features with no qrels leakage---still produces
15.0\% bait. LLM CER, using semantically correct conceptual labels,
produces 29.2\%. Better supervision reduces bait but does not eliminate
it. Second, IDF filtering makes things worse: the filtered LLM CER
variant has a higher bait rate (31.2\%) and top-1 bait rate (18.5\%)
than the unfiltered version, while also reducing RelCov from 97.1\% to
91.2\%. Frequency and discrimination are not the same thing. Third, the binary-derived row reveals the structural origin of the problem: near-zero bait
is achievable---but only by accepting RelCov of 19.7\%.
Discriminativeness and coverage are in direct tension because entities
specific enough to be highly discriminative are rare enough to be absent
from most relevant documents.

\subsection{Why Supervision Cannot Escape the Dilemma}
\label{subsec:Binary Derivation}

The bait rates above are properties of trained rankers. But the
supervision noise is already present before any model training begins,
in the binary derivation rule itself.

Binary derivation partitions entities per query into exclusive positives
($\mathcal{E}^+$, labeled relevant), exclusive negatives ($\mathcal{E}^-$,
labeled irrelevant), and common ($\mathcal{E}^0$, discarded). The three
partitions have a 51:1 imbalance of exclusive negatives to positives
(11,436,791 vs.\ 240,138 entity-query pairs), meaning supervision is
dominated by negative evidence from the outset. Common entities are
generic on aggregate statistics (mean IDF 6.23, median DF 1,198---
comparable to semantic stopwords), which is what motivates their
exclusion. As we show below, however, frequency is a poor proxy for
discriminative value, and these aggregate statistics conceal a
discriminative majority within the partition. Exclusive positives and
negatives are both rare (mean IDF
9.87 and 9.00, median DF 27 and 69 respectively), reflecting the rarity
bias the derivation rule introduces. But the more important question is
what filtering discards.

\begin{figure}[t]
\centering
\includegraphics[width=0.8\columnwidth]{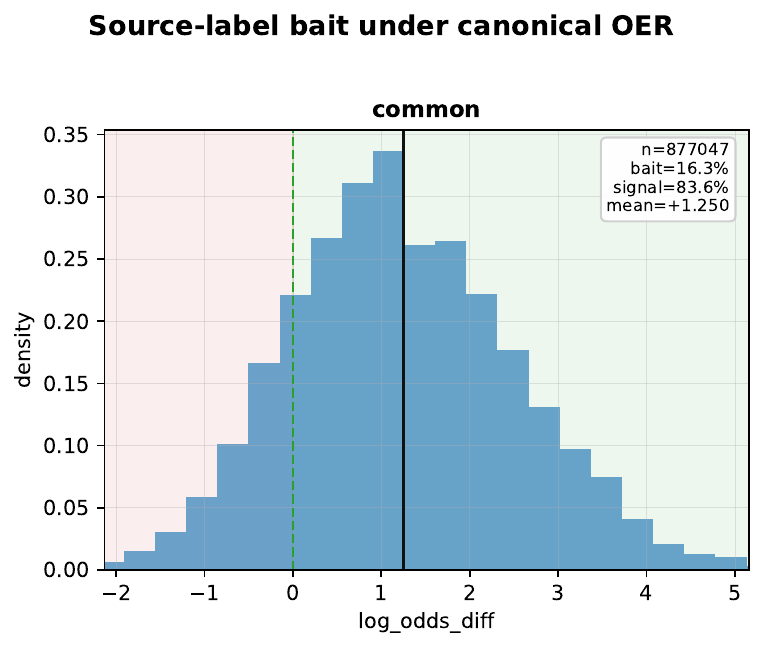}
\caption{OER log-odds distribution of the common entity partition
($n = 877{,}047$ pairs discarded by binary derivation). Mean $= +1.25$
(vertical line); zero marked with dashed line. 83.6\% of discarded
entities are discriminative under OER; only 16.3\% are bait. The
distribution is centred well into the signal region, confirming that the
common partition is a mixed pool rather than a noise pool.}
\label{fig:common_logodds}
\end{figure}

Applying canonical OER scores to the common partition directly: 83.6\%
of discarded entities have positive log-odds differences (mean $= +1.25$,
median $= +1.18$) --- they appear more frequently in relevant than
non-relevant documents and are discriminative by OER criteria, just not
exclusively so. Figure~\ref{fig:common_logodds} shows that the common
partition is a mixed pool, not a noise pool, centred well into the signal
region. Binary filtering correctly removes the 16.3\% bait tail but
discards the 83.6\% discriminative majority alongside it. The
exclusive-presence criterion captures rarity, not utility.

Binary filtering therefore achieves adequate precision for bait exclusion
but poor recall for signal inclusion: it discards 83.6\% of
discriminative entity-query pairs to eliminate the 16.3\% bait tail.
The exclusive-presence criterion ($df_{\text{nonrel}} = 0$) is
correlated with rarity, not with discriminative utility. The result is a supervised entity ranker trained on labels that
systematically favor low-frequency, idiosyncratic entities over
moderate-frequency discriminative ones. This is the same structural bias
that produces the coverage ceiling in

Section~\ref{subsec:The Coverage Ceiling}.

The corruption is not introduced by model training; it is inherited from the
source labels.

\subsection{Why Training Compounds the Problem}
\label{subsec:Training Signal Corruption}

Even within the entity pool that binary derivation retains, the training
signal is further corrupted by hard negative mining. With mean IDF 10.73
and median DF 7, hard negatives are substantially rarer and more
query-specific than random negatives (IDF 8.86, DF 84). The ranker
learns to separate informative entities from semantically plausible, rare
alternatives, which makes the task meaningful but unstable. Across 249
queries, 48.5\% of hard negatives appear in the relevant sets of other
queries (std.\ 25.2\%), compared with 84.2\% for random sampling ---
hard negative mining helps, but nearly half the negative training signal
remains contaminated. At the per-query extreme, some queries have a
cross-query rate of 1.0. The fundamental problem is the entity space
itself: in a news collection like Robust04, topically related entities
recur across queries, making it structurally difficult to find negatives
that are globally irrelevant.

The rarity bias compounds this. Because hard negatives are skewed toward
rare entities (median DF 7), the model is exposed to a negative
distribution biased toward rare, plausible alternatives that happen to
be relevant elsewhere in the collection. A ranker that learns to treat
these as negatives may do so by learning surface sparsity rather than
genuine discriminative features --- the same rarity signal that binary
derivation rewards in the positive set. Supervision noise and training
corruption are therefore not independent failure modes; they share a
common cause: the entity linking environment on Robust04 makes it
structurally difficult to construct a training signal that is both clean
and generalizable.

\subsection{Synthesis}
\label{subsec:4_Synthesis}

The three preceding subsections establish the mechanism behind the
dilemma stated at the opening of this section. The coverage ceiling
(Section~\ref{subsec:The Coverage Ceiling}) shows that entity signals
cannot simultaneously reach most relevant documents and remain
discriminative --- expanding entity selection always buys coverage at
the cost of discrimination, at every value of $k$ tested. Binary
derivation (Section~\ref{subsec:Binary Derivation}) shows that this
tradeoff originates at the source: the supervision rule rewards rarity,
not utility, and discards 83.6\% of the discriminative signal it could
otherwise provide. Training corruption
(Section~\ref{subsec:Training Signal Corruption}) shows that even within
the retained pool, nearly half the negative training signal is
contaminated by cross-query false negatives --- a structural consequence
of entity relevance being entangled across queries in a topically diverse
collection.

These failure modes reinforce each other. The entity selections that are
clean enough to discriminate are too sparse to cover most relevant
documents. The selections broad enough to cover relevant documents are
too noisy to discriminate. And the supervision available to resolve this
tension is itself corrupted in a way that cannot be fixed within the
current benchmark construction. The result is the structural
\emph{supervision--coverage dilemma}: there is no configuration that
simultaneously provides the document reachability needed for effectiveness under clean supervision and the entity precision needed for meaningful discrimination. Crucially, this ceiling is set by the benchmark environment---the
linking incompleteness and the derivation rule---not by model quality.

This diagnosis raises an immediate question: can the dilemma be
mitigated by explicitly filtering entity signals toward more
discriminative ones? Section~\ref{sec:Empirical Evidence for the
CER--OER Gap} tests this directly and shows that post-hoc OER filtering
improves discriminative quality but only by reducing coverage further,
leaving ranking performance below BM25 at every threshold. The dilemma
is not resoluble by signal cleaning alone.

The next section provides the conceptual framework for understanding
precisely why---by separating what entity signals mean semantically
from what they provide observably in the linked collection.

\section{Observable vs.\ Conceptual Entity Relevance}
\label{sec:Observable vs Conceptual Entity Relevance}

Section~\ref{sec:The Supervision--Coverage Dilemma} established that the
supervision--coverage dilemma is structural. But it left a question
unanswered: why does improving supervision quality not help? The LLM CER
labels are semantically correct by construction, yet 29.2\% of its
top-20 selections are bait and its rankings are only weakly aligned with
observably discriminative entities (Spearman $= +0.131$). The
binary-derived ranker is \emph{negatively} aligned (Spearman $= -0.210$).
These are not model failures. They are symptoms of a mismatch between
what supervision measures and what retrieval requires.

The distinction formalised in Section~\ref{subsec:Observable Entity
Relevance} makes this precise. \emph{Conceptual Entity Relevance} (CER)
captures whether an entity is semantically related to a query ---
a judgment made from query text and entity descriptions alone, without
reference to the collection or linking environment. \emph{Observable
Entity Relevance} (OER) measures whether an entity's presence in a
document, as detected by the entity linker in this collection, raises
the probability that the document is relevant. CER is a property of the
query--entity pair; OER is a property of the query--entity--collection
triple. No supervision strategy evaluated in this paper targets OER.
Binary derivation targets exclusivity, which is a proxy for rarity; LLM
labels target conceptual relatedness; corpus-grounded selectors
approximate discriminativeness only indirectly, through frequency
statistics. None accounts directly for the linking environment that
determines observable discriminativeness. This is the root cause of
coverage loss, bait selection, and the collapse under clean supervision.

\subsection{Three Ways CER-Based Supervision Diverges from OER}
\label{subsec:Empirical Evidence and Failure Modes}

\begin{figure}[t]
\centering
\includegraphics[width=0.8\columnwidth]{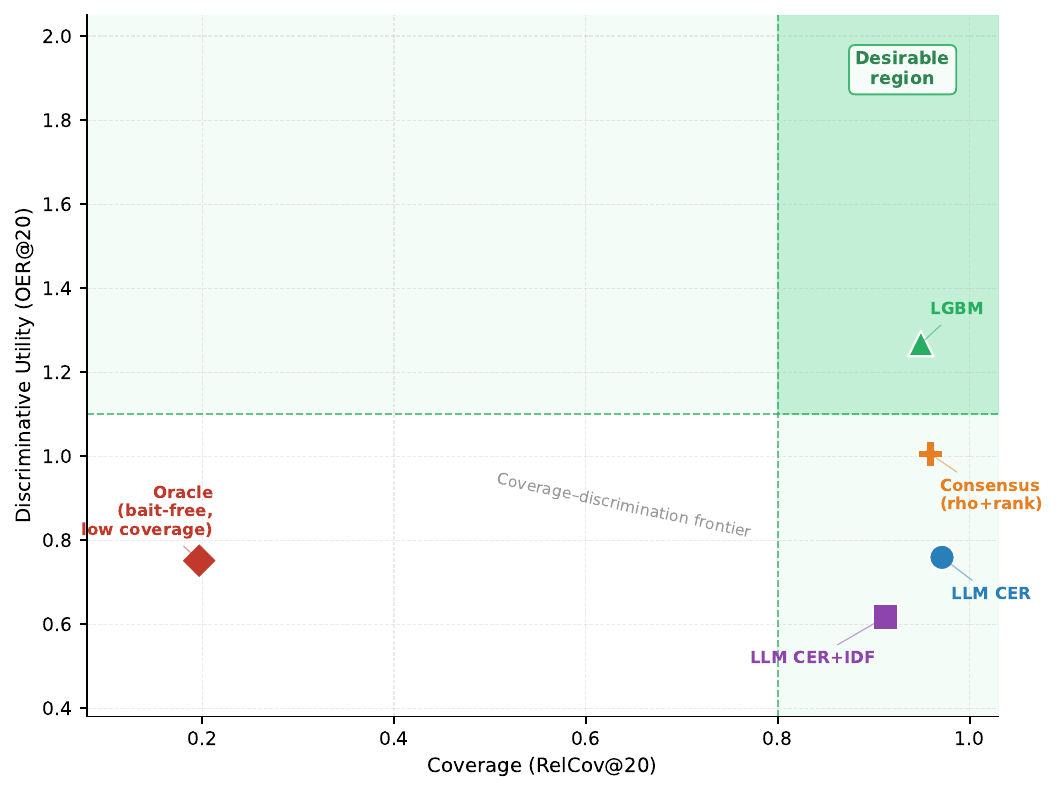}
\caption{Coverage--discrimination space for all evaluated configurations.
Horizontal axis: RelCov@20 (fraction of relevant documents reachable
through the entity channel). Vertical axis: OER@20 (mean log-odds
discriminativeness of selected entities). The desirable region (upper
right, shaded) requires both simultaneously. No leakage-free
configuration occupies it. Arrows trace the trade-off from binary-derived
(bait-free, low coverage) through LLM CER to LGBM (best leakage-free
compromise).}
\label{fig:relcov_oer}
\end{figure}

Table~\ref{tab:cer_oer} quantifies CER--OER alignment for each
configuration: mean OER log-odds of top-20 selected entities (OER@20),
bait and top-1 bait rates, Spearman correlation between ranker scores
and OER log-odds, and exact relevant-document coverage (RelCov@20, from
Table~\ref{tab:coverage}). Figure~\ref{fig:relcov_oer} plots each
configuration in coverage--discrimination space. No leakage-free
configuration occupies the desirable upper-right quadrant --- high
coverage and high discriminative utility simultaneously. The single
configuration that approaches it selects entities directly by their
measured OER on the test judgments, reaching RelCov@20 $= 0.728$ at
$k = 20$ and $0.871$ at $k = 50$; it is an oracle diagnostic rather than
a deployable method, and its position is the clearest evidence that the
quadrant is empty for want of annotation rather than as a matter of
principle. The results reveal three distinct failure modes.

\begin{table}[t]
\centering
\caption{CER vs.\ OER alignment across the supervision ladder at $k=20$.
OER@20 is the mean log-odds discriminativeness of selected entities;
Signal\% is the fraction classified as core or conditional signal;
Spearman is the correlation between ranker scores and OER log-odds.
RelCov@20 is from Table~\ref{tab:coverage}. The binary-derived run is negatively
aligned with OER despite near-zero bait, revealing that bait-free
selection is not equivalent to observably discriminative selection.}
\label{tab:cer_oer}
\begin{tabular}{lrrrrrr}
\toprule
\textbf{Run} & \textbf{OER@20} & \textbf{Bait\%} & \textbf{Signal\%} &
\textbf{Top-1 Bait\%} & \textbf{Spearman} & \textbf{RelCov@20} \\
\midrule
BERT (binary-derived, qrels) & 0.75 & 0.2  &  7.0 &  0.0 & $-$0.210 & 0.197 \\
BERT (LLM CER)                 & 0.76 & 29.2 & 67.0 &  9.6 & $+$0.131 & 0.971 \\
BERT (LLM CER + IDF)           & 0.62 & 31.2 & 59.0 & 18.5 & $+$0.137 & 0.912 \\
LGBM selector                  & 1.27 & 15.0 & 79.0 &  8.0 & $+$0.405 & 0.949 \\
Consensus (rho+rank)           & 1.01 & 21.3 & 73.0 & 19.6 & $+$0.061 & 0.959 \\
\bottomrule
\end{tabular}
\end{table}

\begin{figure}[t]
\centering
\includegraphics[width=\columnwidth]{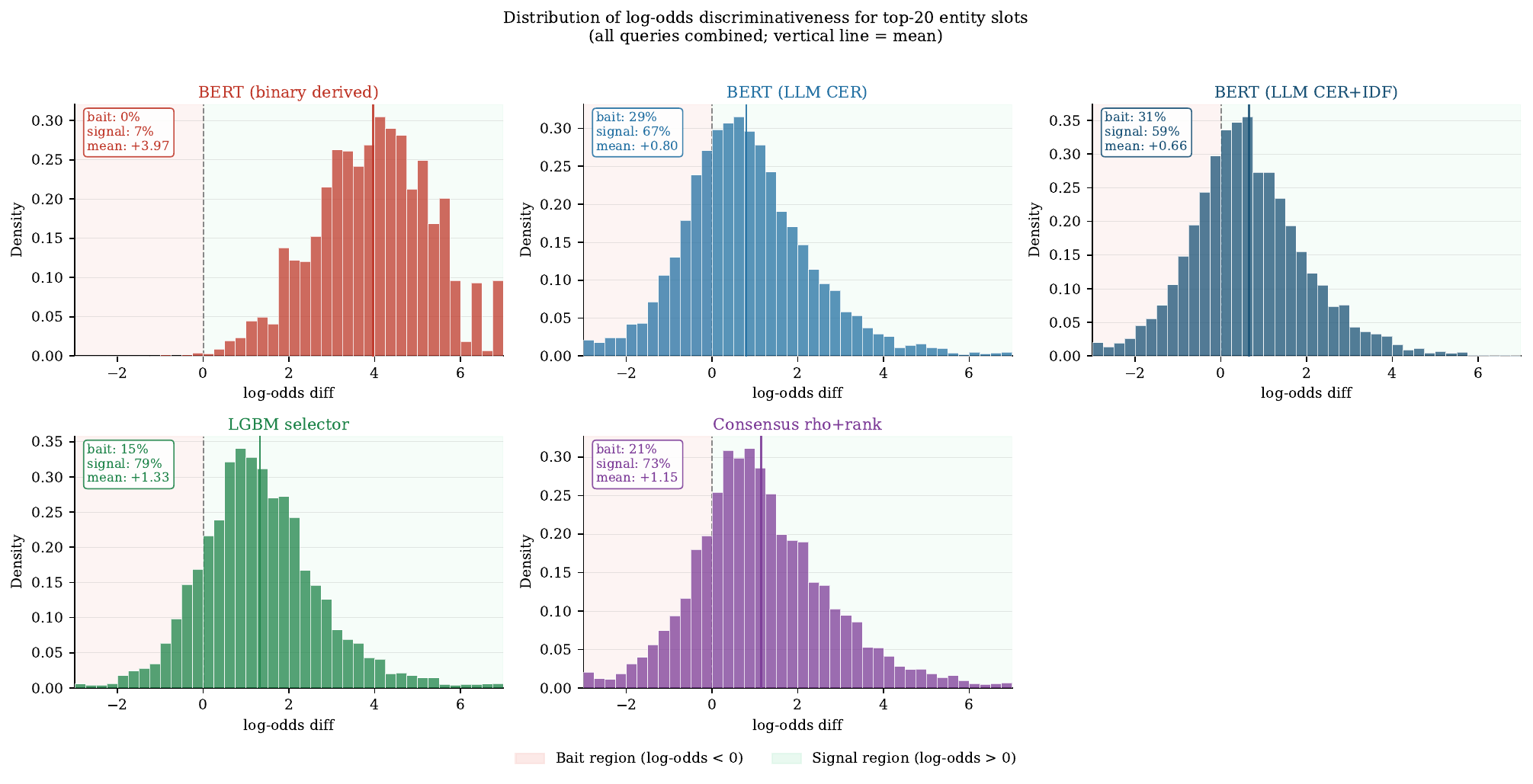}
\caption{Distribution of OER log-odds discriminativeness for top-20
entity slots across all queries, per configuration. The binary-derived panel
(top left) shows near-zero bait but only 7\% signal --- 93\% falls into
the sparse-evidence category. The LLM CER panel shows a broad bait tail
(29\%). The LGBM panel achieves the best shift toward positive log-odds
(mean $+1.33$) with the smallest bait tail (15\%). All panels confirm
the same structural constraint: the linking environment bounds achievable
discriminative quality regardless of supervision strategy.}
\label{fig:logodds_distributions}
\end{figure}

\medskip
\noindent
\textbf{Failure mode 1: Binary derivation is bait-free but
observationally sparse.}
The binary-derived run achieves Bait\% $= 0.2$ and Top-1 Bait\% $= 0.0$,
yet its Spearman with OER is $-0.210$ and Signal\% is only 7.0 --- 93\%
of its top-20 entity slots fall into the sparse-evidence category
(Figure~\ref{fig:logodds_distributions}, top left). The explanation
connects directly to Section~\ref{subsec:Binary Derivation}: binary
derivation rewards entities appearing exclusively in relevant documents
($df_{\text{nonrel}} = 0$), which are by definition rare. OER weights
log-odds by candidate-pool frequency, penalising rare entities regardless
of directional sign. The same rarity bias that causes binary derivation
to discard 83.6\% of discriminative entities at the source also causes
the trained ranker to select entities too sparse to score as OER signal.
Bait-free is not the same as observably discriminative.

\medskip
\noindent
\textbf{Failure mode 2: LLM CER selects plausible but
non-discriminative entities.}
The LLM CER run achieves Spearman $= +0.131$ with Bait\% $= 29.2$ and
Signal\% $= 67.0$. Nearly a third of its top-20 selections are
non-discriminative despite being generated by a model reasoning
explicitly about topical relevance. The LLM cannot know which entities
WAT consistently annotates in Robust04, nor which appear at similar
rates across relevant and non-relevant documents in the candidate pool.
It makes correct CER judgments that are disconnected from the observable
linking environment. IDF filtering does not fix this: LLM CER+IDF has a
higher bait rate (31.2\%), higher top-1 bait (18.5\%), and lower coverage
(0.912), with Spearman barely changed ($+0.137$). No post-hoc
reweighting resolves a mismatch between conceptual relevance and
observable discriminative utility.

\medskip
\noindent
\textbf{Failure mode 3: Corpus-grounded selection achieves the best
compromise but does not escape.}
The LGBM selector achieves the highest OER@20 (1.27), lowest bait rate
(15.0\%), highest signal rate (79.0\%), and highest Spearman ($+0.405$),
without any qrels-derived features --- using only corpus statistics,
BM25 rank, and lexical overlap. The consensus (rho+rank) variant reaches
a similar position (OER@20 $= 1.01$, Bait\% $= 21.3\%$) without
training. Both remain on the frontier rather than escaping it
(Figure~\ref{fig:relcov_oer}): the best available proxy for OER still
yields 15\% bait and moderate Spearman. The entity channel is bounded
not by semantic model quality but by the linking environment itself.

\subsection{Implications}
\label{subsec:Implications}

\medskip
\noindent
\textbf{The collapse under clean supervision is the CER--OER gap at
scale.}
Every entity-oriented model collapses to BM25-level performance once
entity selection must be performed without relevance judgments, because
no available supervision targets OER while ranking depends on it. Under
leaked supervision the entity run encodes the judgments directly, so the
gap is invisible; withdrawing them exposes two consequences at once.
Entities selected for exclusivity fail to reach most relevant documents
(coverage ceiling); entities selected for conceptual relatedness fail to
distinguish relevant from non-relevant ones (discrimination floor). The
CER--OER gap is not a bug in any specific model. It is a consequence of
building entity-oriented systems on collections that annotate neither
quantity.

\medskip
\noindent
\textbf{Corpus-grounded supervision is the productive direction.}
Closing the gap requires collection-grounded signals: entity frequency,
candidate-pool prevalence, and the statistical relationship between
entity presence and document relevance. The LGBM selector's implicit
approximation of this through corpus statistics is the closest any
evaluated method comes to targeting OER directly. Supervision strategies
that explicitly target OER rather than CER would require
collection-specific annotation infrastructure beyond what standard
benchmarks currently provide.

\medskip
\noindent
\textbf{Evaluation should separately measure CER and OER.}
Current benchmarks evaluate document ranking quality but do not measure
whether entity selection is operating at the CER or OER level. The
leaked vs.\ clean supervision contrast surfaces this at the document level.
The OER@20, Signal\%, and Spearman metrics in Table~\ref{tab:cer_oer}
provide complementary entity-level diagnostics that reveal \emph{why} a
model's entity channel succeeds or fails, rather than only whether the
final ranking is correct.

\section{Empirical Evidence for the CER--OER Gap}
\label{sec:Empirical Evidence for the CER--OER Gap}

Section~\ref{sec:Observable vs Conceptual Entity Relevance} argued that
the supervision--coverage dilemma persists because no available
supervision targets OER while retrieval depends on it. This section provides three lines of
empirical support for that claim. We show that performance gains appear
only where coverage is discriminative rather than merely broad; that no
configuration across 193 entity selection strategies escapes the
coverage--discrimination frontier; and that explicitly enforcing OER
alignment improves signal quality but reduces coverage to the point that
ranking falls below BM25. Once the system is analyzed through an OER
lens, the same conclusion appears at every level: only discriminative
coverage predicts gains, the global run space remains trapped on the same
structural frontier, and directly cleaning the signal cannot overcome
coverage loss.

This pattern arises before model training begins. Binary derivation
discards entities that appear in both relevant and non-relevant
documents, using exclusivity as a proxy for discriminative value. But
exclusivity captures rarity, not utility: 83.6\% of discarded entities
are in fact discriminative under OER. At the same time, rare but
non-informative entities are retained. The cross-query false negative
problem has the same source: neither exclusivity nor conceptual
relatedness conditions on whether an entity is discriminative in the
candidate pool for this query in this collection. As a result, both
training noise and bait selection arise from the same structural
mismatch. The following analyses show how this
mismatch manifests in query-level gains, the global configuration space,
and direct filtering interventions.

\subsection{Only Observable Coverage Predicts Gains}
\label{subsec:Coverage Controls Performance Gains}

Partitioning the 249 Robust04 queries into tertiles by RelCov@20 of the
LGBM reference run confirms directionally that entity model gains
increase with coverage: the best gain in the high-coverage bucket reaches
$\Delta$MAP $= +0.031$. But BM25 MAP itself rises by 0.105 across the
same buckets --- a gap that dwarfs entity model gains and reflects that
high-coverage queries are topically focused queries term-based retrieval
already handles well. When each model's queries are bucketed by its own
RelCov@20, BERT (LLM CER+IDF) falls \emph{below} BM25 in the
low-coverage bucket (MAP 0.234 vs.\ 0.251), showing that reducing
coverage without recovering discrimination actively harms performance.
The gradient therefore reflects query difficulty as much as entity-channel
quality.

Table~\ref{tab:exp4} tests the coverage--gain relationship with OLS
regression, controlling for query difficulty (BM25 AP and number of
relevant documents), across four coverage measures. The divergence across
measures is the main finding.

\textbf{Reference-run coverage} (RelCov@20 of the shared LGBM run)
produces near-zero $R^2$ (0.001--0.023) and insignificant continuous
coefficients for all models. A breakpoint sweep identifies a consistent
structural discontinuity at RelCov $= 0.980$ across all four models:
below this threshold (90 queries), mean $\Delta$AP is 0.011--0.017;
above it (158 queries), gains roughly double to 0.018--0.031, with the
difference significant for BERT LLM CER ($p = 0.044$) and marginally
significant for LGBM ($p = 0.063$). This suggests a minimum coverage
floor below which the entity channel contributes little --- but not a
monotone relationship above it.

\textbf{Model-own coverage} (each model's own RelCov@20) reverses the
sign: coefficients are uniformly negative. Within the high-coverage
regime all practical models already occupy (mean own-RelCov $> 0.95$),
additional coverage is associated with more generic bait and lower
discrimination. More coverage from a noisier entity signal hurts.

\textbf{Oracle coverage} (fraction of relevant documents reachable by a
greedy minimum-cover oracle; mean 0.999, 231 of 248 queries at 1.0)
produces negative coefficients for all models, significant for BERT LLM
CER ($\hat{\beta} = -0.644$, $p = 0.083$) and Consensus
($\hat{\beta} = -0.598$, $p = 0.029$). Queries where perfect coverage is
achievable are precisely those where BM25 is already strong. Coverage
availability does not translate into gains when the underlying queries
are easy.

\textbf{Observable coverage} (fraction of relevant documents reachable
through OER label-2 core-signal entities) is the exception: coefficients
are positive for all models --- the only coverage measure with a
consistent positive direction --- and reach significance for BERT LLM
CER+IDF ($\hat{\beta} = 0.039$, $p = 0.034$). Coverage through
discriminative entities predicts gains; coverage through all entities
does not. This is the CER--OER gap in operational form: the distinction
between conceptual and observable relevance is directly visible in which
coverage measure predicts system behavior.

\begin{table*}[t]
\centering
\caption{Regression analysis of per-query $\Delta$AP on four coverage
measures, controlling for BM25 AP and number of relevant documents.
OLS $R^2$ and $\hat{\beta}$ (continuous coverage coefficient with sign)
are shown for each model and coverage measure. Breakpoint $\tau$ and its
$p$-value are from the two-segment OLS sweep (reference-run coverage
only). Observable coverage is the only measure with a consistent
positive direction across all models. Significance: $^{***}p<0.001$,
$^{**}p<0.01$, $^{*}p<0.05$, $^{\dagger}p<0.10$.}
\label{tab:exp4}
\begin{tabular}{lcccccccccc}
\toprule
\textbf{Model} &
\multicolumn{2}{c}{\textbf{Ref.\ coverage}} &
\multicolumn{2}{c}{\textbf{Own coverage}} &
\multicolumn{2}{c}{\textbf{Oracle coverage}} &
\multicolumn{2}{c}{\textbf{Observable coverage}} &
\multicolumn{2}{c}{\textbf{Breakpoint}} \\
\cmidrule(lr){2-3}\cmidrule(lr){4-5}\cmidrule(lr){6-7}
\cmidrule(lr){8-9}\cmidrule(lr){10-11}
 & $R^2$ & $\hat{\beta}$ & $R^2$ & $\hat{\beta}$ &
   $R^2$ & $\hat{\beta}$ & $R^2$ & $\hat{\beta}$ &
   $\tau$ & $p$ \\
\midrule
BERT (LLM CER) &
  0.023 & $+0.039$ & 0.019 & $-0.016$ &
  0.025 & $-0.644^{\dagger}$ & 0.025 & $+0.025$ &
  0.980 & $0.044^{*}$ \\
BERT (LLM CER+IDF) &
  0.013 & $+0.031$ & 0.010 & $-0.009$ &
  0.011 & $-0.219$ & 0.026 & $+0.039^{*}$ &
  0.980 & 0.126 \\
LGBM selector &
  0.001 & $+0.017$ & 0.003 & $-0.048$ &
  0.002 & $-0.410$ & 0.002 & $+0.013$ &
  0.980 & $0.063^{\dagger}$ \\
Consensus rho+rank &
  0.015 & $+0.031$ & 0.023 & $-0.097$ &
  0.017 & $-0.598^{*}$ & 0.018 & $+0.026$ &
  0.980 & 0.239 \\
\bottomrule
\end{tabular}
\end{table*}

The four measures converge on one conclusion. A minimum coverage floor
exists at RelCov $\approx 0.98$ below which the entity channel is too
sparse to contribute. Above that floor, what matters is not how many
relevant documents the channel reaches but whether it reaches them
through discriminative entities. Raw, oracle, and model-own coverage all
show zero or negative relationships to gains once query difficulty is
controlled. Only observable coverage has a consistent positive direction.

\subsection{The Coverage--Discrimination Frontier Is Global}
\label{subsec:The Coverage--Discrimination Frontier Is Structural}

\begin{figure}[t]
    \centering
    \includegraphics[width=\linewidth]{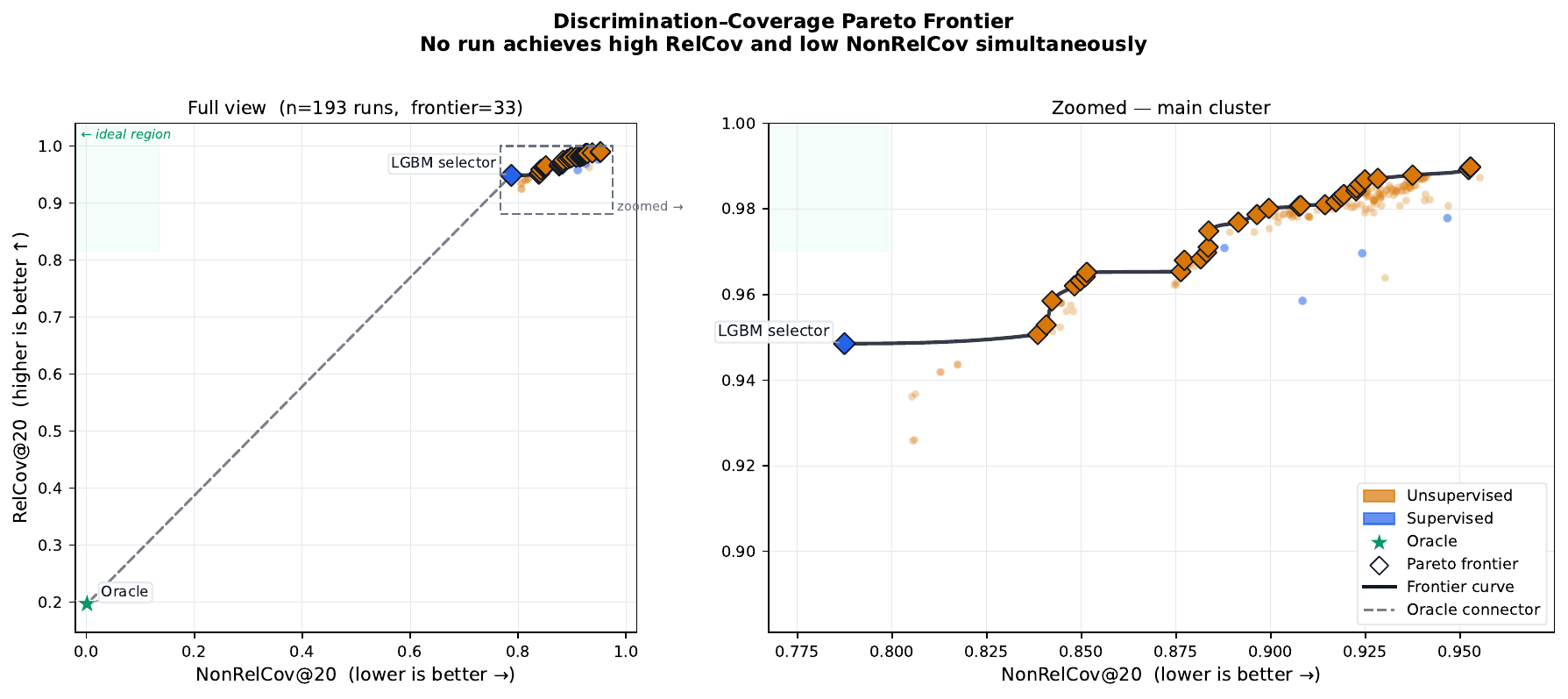}
    \caption{Discrimination--coverage Pareto frontier for all 193 entity
    selection configurations on Robust04 (6 supervised, 187 unsupervised).
    Horizontal axis: NonRelCov@20 (lower is better); vertical axis:
    RelCov@20 (higher is better). The green shaded region marks the
    desirable operating regime (RelCov $> 0.80$, NonRelCov $< 0.50$).
    Panel~(a): full range including the binary-derived run (lower-left star).
    Panel~(b): zoomed main cluster with Pareto frontier curve. No
    leakage-free configuration occupies the desirable region at any threshold.}
    \label{fig:pareto_frontier}
\end{figure}

The coverage--discrimination tradeoff established in
Section~\ref{subsec:The Coverage Ceiling} for five representative runs
holds across the entire configuration space.
Figure~\ref{fig:pareto_frontier} plots all 193 entity selection
configurations --- six supervised and 187 unsupervised spanning six
method families --- in the NonRelCov@20 vs.\ RelCov@20 plane.

The ideal operating region (RelCov $> 0.80$, NonRelCov $< 0.50$) is empty
of leakage-free configurations at every threshold examined. No
leakage-free run satisfies even the relaxed criteria of RelCov $> 0.90$
and NonRelCov $< 0.70$, nor RelCov $> 0.95$ and NonRelCov $< 0.75$.
Configurations that use the test judgments do occupy the region---entity
selection driven directly by measured OER reaches RelCov $= 0.871$ at
NonRelCov $= 0.258$ for $k = 50$---which locates the emptiness in the
absence of OER annotation rather than in any property of the entity
channel itself. Among the 193 leakage-free runs, RelCov@20
ranges from 0.926 to 0.990 (mean 0.977) and NonRelCov@20 from 0.787 to
0.955 (mean 0.908). The maximum DiscRatio achieved by any leakage-free run
is 1.204 (LGBM), and the mean coverage gap (RelCov $-$ NonRelCov) is
only 0.069. Pearson $r = 0.956$ between RelCov and NonRelCov across all
193 runs confirms that coverage and contamination are structurally
coupled: every run that reaches more relevant documents inevitably reaches
more non-relevant ones at nearly the same rate.

The frontier's composition is as informative as its shape. Of the 32 leakage-free frontier runs, 31 are unsupervised and only one (LGBM) is supervised. Supervised training does not push models into a
meaningfully better region of the coverage--discrimination space. The
tradeoff is not a failure of supervision; it is a structural property of
the collection and linking environment that no amount of learning
resolves.

The $k$-progression confirms the frontier cannot be escaped by selecting
more entities. For LGBM, RelCov rises from 85.6\% to 99.6\% as $k$ goes
from 10 to 50, while DiscRatio falls from 1.44 to 1.03. For the binary-derived run,
RelCov rises from 11.8\% to 33.8\% while DiscRatio falls from 211.9 to
96.4. The frontier is a slope, not a point: every increment of coverage
costs a corresponding decrement in discrimination, at every value of $k$,
for every configuration tested.

The document-ranking view confirms the same pattern. Even among the 102
of 248 queries (41.1\%) where core-signal OER entities already cover
every relevant document in the candidate pool (ObsCov $= 1.0$), entity
models do not achieve large gains over BM25. Perfect observable coverage
is therefore not sufficient: when the linking environment limits
discrimination, additional coverage cannot translate into improved
ranking.
\subsection{OER Filtering Confirms the Dilemma Is Inescapable}
\label{subsec:Attempting to Bridge the CER--OER Gap via OER Filtering}

The two preceding subsections establish the tradeoff from the coverage
side. A direct test of whether it can be escaped is to explicitly filter
entity signals toward OER-aligned selections and observe the effect on
document ranking. We apply OER thresholds $\tau \in \{0.0, 0.5, 1.0\}$
to the LLM CER entity run (RelCov@20 $= 0.971$, DiscRatio@20 $= 1.09$,
MAP $= 0.310$) before passing entities to the document ranker, holding
the ranking model fixed throughout. This isolates the effect of entity
signal quality from any model retraining effect.

Table~\ref{tab:block6} shows the results. OER filtering works as
intended at the entity level: DiscRatio@20 rises from 1.219 at $\tau=0.0$
to 1.769 at $\tau=1.0$, and NonRelCov@20 falls from 0.735 to 0.373. But
the improvement comes at a monotonically increasing cost in coverage:
RelCov@20 drops from 0.895 at $\tau=0.0$ to 0.660 at $\tau=1.0$,
leaving 34\% of relevant documents structurally unreachable at the
strictest threshold.

\begin{table}[t]
\centering
\small
\caption{OER filtering results. Starting from the unfiltered LLM CER
entity run, OER thresholds $\tau \in \{0.0, 0.5, 1.0\}$ are applied
before passing entities to the document ranker (model weights held
fixed). As $\tau$ increases, discrimination improves but
relevant-document coverage drops, and document ranking falls below both
the unfiltered baseline and BM25+RM3 at every threshold. \textbf{The
filtered rows apply OER thresholds computed from the test relevance
judgments and are oracle diagnostics, not deployable configurations};
they bound what OER-aligned filtering could achieve, and that bound is
below BM25. Note also that $\tau = 0.0$ is not the unfiltered run: it
already removes entities with non-positive OER, reducing RelCov from
0.971 to 0.895.}
\label{tab:block6}
\begin{tabular}{lcccccc}
\toprule
\textbf{Run} & \textbf{RelCov@20} & \textbf{NonRelCov@20} &
\textbf{DiscRatio@20} & \textbf{MAP} & \textbf{nDCG@20} & \textbf{P@20} \\
\midrule
BM25+RM3 (baseline)     & ---   & ---   & ---   & 0.292 & 0.435 & 0.383 \\
LLM CER (unfiltered)    & 0.971 & 0.888 & 1.09  & 0.310 & 0.464 & 0.404 \\
\midrule
OER filtered $\tau=0.0$ & 0.895 & 0.735 & 1.219 & 0.210 & 0.352 & 0.307 \\
OER filtered $\tau=0.5$ & 0.797 & 0.560 & 1.423 & 0.221 & 0.373 & 0.328 \\
OER filtered $\tau=1.0$ & 0.660 & 0.373 & 1.769 & 0.232 & 0.386 & 0.337 \\
\bottomrule
\end{tabular}
\end{table}

Document ranking does not recover. Filtered runs achieve MAP of 0.210,
0.221, and 0.232 --- all substantially below the unfiltered baseline
(0.310) and BM25 (0.292). The monotonic rise from $\tau=0.0$ to
$\tau=1.0$ reflects partial recovery of discrimination, but even the
strictest threshold leaves ranking far below the unfiltered level. The
same pattern holds for nDCG@20 (0.352, 0.373, 0.386 versus 0.464) and
P@20 (0.307, 0.328, 0.337 versus 0.404). Every filtered run falls below
BM25 because coverage loss dominates: once stripped of its
broad-coverage entities, the entity channel provides less useful signal
than a well-tuned term-based retriever. OER can clean the signal, but it
cannot recover relevant documents the entity linker never annotated. The
CER--OER gap is not a conceptual story. It is a measurable, structural
property of entity-oriented retrieval on standard ad hoc collections,
visible in query-level performance patterns, in the global configuration
space, and in direct filtering interventions alike.

\section{Implications}
\label{sec:Implications}

The introduction asked whether current evaluation settings in
entity-oriented retrieval can meaningfully assess the usefulness of entity
signals for document ranking. The answer this paper's evidence supports is
no---but for a precise reason. Evaluation under leaked entity
supervision cannot, because the entity run encodes the judgments the
ranking is scored against, so effectiveness reflects the supervision
rather than the model. Evaluation under clean supervision cannot, because
it is dominated by entity linking coverage rather than model quality,
making a system that works well in principle indistinguishable from one
that fails entirely. Both regimes fail to
isolate what they purport to measure, and reporting one as if it implied
the other is a category mistake that has shaped how the field interprets
entity-oriented results. The three implications below follow from this
diagnosis.

\subsection{Leaked and Clean Entity Supervision Are Answering Different
Scientific Questions}
\label{subsec:Conditional and Open-World Evaluation Are Not Interchangeable}

Leaked supervision asks: \textit{can the model exploit entity evidence
effectively when that evidence is aligned with relevance?} Clean
supervision asks: \textit{does the full entity-oriented pipeline improve
retrieval when that alignment must be inferred from the collection?}
These are not variants of the same question. They are different
scientific questions, and they admit different answers for the same model.

The mechanism is the entity run, not the evaluation pool. Both figures for
QDER are computed over the full 1000-document candidate set with
unrestricted recall denominators; the 80.3\% of relevant documents that
binary-derived selection cannot reach are counted against the model in
both cases. What changes is whether the entity run itself was constructed
from the judgments. A model can therefore achieve MAP 0.608 under leaked
supervision and 0.294 under clean supervision simultaneously---as the
binary-derived QDER run does---because the entities the judgments
identify are highly discriminative but rare. Conversely, a model can show
almost no gap---as the LLM CER runs do---not because it succeeds, but
because its entities are broad enough to be selectable without judgments
at all. In both cases the coverage ceiling is the same. What changes is
which failure mode is visible.

Across 443 evaluated configurations, gains reported under leaked
supervision do not carry over once entity selection is made independent
of the judgments. The best improvement over BM25 under clean supervision
is 0.051 MAP points. This is not
evidence that entity-oriented architectures are weak---the best
open-world configuration matches the TREC Robust04 best system and
outperforms the majority of neural rerankers. It is evidence that the
two evaluation regimes are measuring different things, and that the field
has not consistently reported both.

The deployment implication is direct. A live retrieval system cannot
derive its entity supervision from the relevance judgments of the queries
it is serving. Clean supervision is the only regime that corresponds to
actual deployment conditions. Yet on current benchmarks it cannot discriminate
between a system that works well in principle and one that fails
entirely, because the signal is dominated by linking coverage rather than
model capacity. The field is therefore in a position where the only
deployable regime is scientifically uninformative, and the only
scientifically informative regime is not deployable. This is not an
argument for choosing one regime. It is an argument for building the
infrastructure that would make open-world evaluation informative, which
we discuss in Section~\ref{subsec:impl_infrastructure}. Until then, both
regimes must be reported---leaked-supervision results for model
capacity, clean-supervision results for end-to-end effectiveness---and
neither should be interpreted as implying the other.

\subsection{Better Entity Supervision Is Not the Path Forward on Current
Benchmarks}
\label{subsec:Derived Entity Supervision Is Structurally Noisy}

The natural response to the CER--OER diagnosis is to seek better entity
supervision. Our results show that on current ad hoc collections, this is
not the limiting factor --- and that refining supervision alone will not
recover open-world performance.

The evidence is direct. Replacing binary-derived labels with LLM-generated
conceptual relevance labels does not resolve the collapse: LLM CER still
produces 29.2\% bait because it assesses topical relevance without access
to the linking environment. A conceptually correct entity label is not
the same as an observably discriminative one, and no amount of semantic
refinement closes that gap. Better LLMs, more annotators, and finer
relevance scales all operate at the CER level: they improve conceptual
relevance judgments, not the observable discriminativeness of selected
entities in the linked collection. The OER filtering experiment shows the
same limitation from the opposite direction: explicitly improving
discriminative quality raises DiscRatio but reduces coverage to the point
that ranking falls below BM25. The two properties supervision would need
to provide --- broad coverage and discriminative precision --- are
structurally opposed under the current linking environment.

The deeper reason is that current benchmarks do not provide an
independent standard for entity relevance. Every supervision strategy is
therefore an imperfect proxy: binary derivation infers entity labels from
document relevance, LLM CER supplies conceptual judgments without
collection grounding, and corpus-based methods approximate OER through
frequency and candidate-pool statistics. None can validate entity
relevance against a benchmark-level annotation standard independent of
document relevance, because no such standard exists. This is why binary
derivation discards 83.6\% of discriminative entities, yields a 51:1
negative-to-positive imbalance, and produces 48.5\% cross-query false
negatives: these are not engineering failures but structural consequences
of deriving entity supervision from document qrels in a topically diverse
collection. Corpus-grounded supervision is therefore the best available
mitigation on current collections---the LGBM selector, trained without
qrels-derived features, achieves the strongest leakage-free tradeoff---but it is not a solution. The solution requires the aligned entity-level
discriminativeness annotations discussed in
Section~\ref{subsec:impl_infrastructure}.

\subsection{Progress Requires Aligned Annotation and Reporting
Infrastructure}
\label{subsec:impl_infrastructure}

The two preceding implications converge on a single conclusion: standard
ad hoc retrieval collections do not provide the annotation infrastructure
that entity-oriented ranking requires. This is not a criticism of any
specific benchmark. Robust04, TREC DL, and similar collections were
designed for document-level retrieval evaluation. The workarounds
currently in use --- binary derivation, entity filtering, conditional
evaluation --- are not approximations that can be refined away. They are
responses to a structural gap between what the collections provide and
what entity-oriented systems need, and they systematically prevent the
field from answering its central question.

The oracle coverage analysis makes the gap concrete. A greedy oracle
selecting the minimum entity set to cover all relevant documents achieves
mean OracleCov@20 $= 0.999$ across 248 queries. The entities required to
cover relevant documents exist in the collection. The bottleneck is not
entity availability but entity identifiability---knowing which of the roughly 31{,}000 entities linked in a query's candidate pool reliably signal relevance rather than
topical adjacency or co-occurrence noise. Selecting twenty such entities
is a needle-in-haystack problem that no available supervision signal is
designed to solve.

Two specific gaps need to be closed. The first is entity-level relevance
annotations. What is needed are collections that pair document relevance
judgments with entity-level discriminativeness annotations for the same
queries---specifying not only which documents are relevant but which
entities reliably distinguish them from non-relevant ones under the
actual linking environment. This would allow entity rankers to be trained
and evaluated directly on OER rather than on the noisy
exclusive-presence proxy that binary derivation provides. TREC Complex
Answer Retrieval~\cite{dietz2017trec} introduced entity annotations
derived from Wikipedia hyperlink structure, but these are not available
for newswire collections and do not provide discriminativeness judgments
grounded in document-level relevance. A parallel extension to
passage-level retrieval would further enable evaluation of
entity-grounded generation pipelines under the same diagnostic framework.

The second gap is evaluation protocol. The leaked vs.\ clean supervision
distinction and the CER--OER framework together motivate a three-level
reporting standard: open-world MAP for end-to-end effectiveness;
entity-level OER metrics (OER@20, Signal\%, Spearman correlation with
OER log-odds) to assess whether selection identifies genuinely
discriminative entities; and coverage-stratified results (RelCov@20
tertiles) to reveal how performance varies with entity-channel reach.
Reporting all three would make explicit what current evaluation conflates---whether gains arise from better entity selection, better document
ranking given entities, or simply from queries where the linking
environment happens to provide adequate coverage.

The constraints themselves expose a structural incompatibility between
standard IR evaluation practice and entity-oriented ranking research. The
field's standard expectation --- demonstrate gains across multiple ad hoc
retrieval collections --- cannot be met fairly without aligned entity and
document relevance annotations that do not exist at scale. Asking
researchers to show entity-oriented gains across collections implicitly
demands something the benchmark environment cannot support, not because
the methods are weak, but because the evaluation infrastructure cannot
provide a fair test. What the field needs, in this light, is a diagnostic
that isolates model capacity---but not the one currently in use.
Supervision derived from the test judgments is not a defensible
diagnostic, because it measures the judgments as much as the model, and
we do not defend it here. A diagnostic is scientifically informative only
if the entity run feeding it is constructed independently of the queries
being evaluated. Our OER-oracle results indicate what such a diagnostic
would show: selection driven by measured discriminativeness reaches
RelCov 0.871 at NonRelCov 0.258, well inside the operating region that no
leakage-free method occupies. The ceiling is therefore a consequence of
missing annotation, not of the entity channel. Building that annotation
would give the field a capacity diagnostic it can defend, and remove any
need for the derived-supervision workaround.

The answer to the question the introduction posed---can current
evaluation settings meaningfully assess the usefulness of entity signals
for document ranking?---is that they can assess one thing each:
evaluation under leaked supervision indicates whether a model exploits
entity evidence when it is aligned with relevance; evaluation under clean
supervision indicates whether entity linking coverage is sufficient for
end-to-end gains. Neither can assess both simultaneously, and neither
should be reported as if it could. Until collections with aligned entity
discriminativeness annotations exist, the field should treat gains under
leaked supervision as an upper bound on model capacity rather than as a
measure of it, clean-supervision results as evidence of pipeline
effectiveness, and the gap between them as a measure of how much of the
entity signal the judgments were supplying---not as a verdict on
whether entity-oriented retrieval works.

\section{Conclusion}
\label{sec:Conclusion}

This paper asked whether current evaluation settings in entity-oriented
retrieval can meaningfully assess the usefulness of entity signals for
document ranking. The answer is precise: they cannot --- not because
entity-oriented architectures are weak, but because the evaluation
environment systematically prevents the question from being answered. The
best configuration under clean supervision achieves MAP 0.343, matching
the TREC Robust04 best system and outperforming the majority of neural
rerankers. The architecture is competitive. What fails is the
infrastructure around it.

The failure is structural and operates on two levels simultaneously.
At the entity-channel level, no leakage-free configuration across 193
entity selection strategies simultaneously achieves the coverage and
discrimination that ranking requires. Entity runs derived from the
judgments are narrow and discriminative but cover only 19.7\% of relevant
documents; runs selectable without judgments cover most relevant documents
but cannot distinguish them from non-relevant ones. The
coverage--discrimination frontier is not a tuning problem---it is a
property of the Robust04 entity linking environment that persists across
all supervised and unsupervised methods, at every value of $k$ tested.
At the supervision level, no strategy available under current benchmark
construction targets OER. Binary derivation targets exclusivity, which is
a proxy for rarity; LLM-generated labels target conceptual relatedness;
corpus-grounded selectors approximate discriminativeness only through
frequency statistics. Better semantic models, more annotators, and finer
relevance scales all improve conceptual judgments. None of them recover
the observable signal the benchmark does not annotate.

The central contribution of this paper is therefore not a new model or a
new benchmark. It is a diagnosis. Leaked and clean entity supervision are
answering different scientific questions --- whether a model can exploit
entity evidence when it is aligned with relevance, and whether the full
pipeline improves retrieval when that alignment must be inferred. Both
questions are valid. Neither subsumes the other. The gains reported in
the entity-oriented literature are arithmetically real, but they are
answers to the first question being read as answers to the second. The
supervision--coverage dilemma and the CER--OER gap together explain why:
the same linking environment that makes results under leaked supervision
look informative makes results under clean supervision uninterpretable,
and the supervision that produces the former is precisely what is
unavailable at deployment.

Entity-oriented retrieval is not a failed research program. It is a
research program that has outgrown the evaluation infrastructure available
to assess it. The path forward is clear: collections with aligned
entity-level discriminativeness annotations, evaluation protocols that
report coverage alongside effectiveness, and a shared understanding that
results under leaked and clean supervision are complementary diagnostics
rather than competing claims about the same thing. Until that
infrastructure exists, the question this paper posed cannot be fairly
answered. What this paper has shown is not that entity-oriented retrieval
cannot work, but what must change before the field can test it fairly.

\bibliographystyle{ACM-Reference-Format}
\bibliography{references}
\end{document}